\documentclass[journal=jpcafh,manuscript=article]{achemso}

\usepackage[T1]{fontenc} 
\usepackage[version=3]{mhchem}
\usepackage{placeins}

\newcommand{\new}[1]{#1}

\newcounter{defcounter}
\newenvironment{reaction}{%
\addtocounter{equation}{-1}
\refstepcounter{defcounter}

\begin{equation}}
{\end{equation}}

\author{April M. Cooper}
\author{Johannes K\"astner}
\affiliation[University of Stuttgart]
{Institute for Theoretical Chemistry,
 University of Stuttgart, 
 Pfaffenwaldring 55, 70569 Stuttgart, Germany}
\email{kaestner@theochem.uni-stuttgart.de}
\phone{+49-711-685-64473}
\title[An \textsf{achemso} demo]
  {Low-Temperature Kinetic Isotope Effects in $\text{CH}_3\text{OH}\,+\,\text{H} \rightarrow \text{CH}_2\text{OH}\,
   +\,\text{H}_2$  Shed Light on the Deuteration of Methanol in Space} 

\begin{document}


\begin{abstract}
  We calculated reaction rate constants including atom tunneling for the
  hydrogen abstraction reaction $\text{CH}_3\text{OH}\,+\,\text{H} \rightarrow
  \text{CH}_2\text{OH}\,+\,\text{H}_2$ with the instanton method.  The
  potential energy was fitted by a neural network, that was trained to
  UCCSD(T)-F12/VTZ-F12 data. Bimolecular gas-phase rate constants were
  calculated using microcanonic instanton theory.  All H/D isotope patterns on
  the CH$_3$ group and the incoming H atom are studied. Unimolecular reaction
  rate constants, representing the reaction on a surface, down to 30~K, are
  presented for all isotope patterns. At 30~K they range from 4100 for
  the replacement of the abstracted H by D to $\sim$8 for the replacement of the
  abstracting H to about 2--6 for secondary KIEs.  The
  $^\text{12}$C/$^\text{13}$C kinetic isotope effect is 1.08 at 30~K, while the
  $^\text{16}$O/$^\text{18}$O kinetic isotope effect is vanishingly small. A
  simple kinetic surface model using these data predicts high abundances of
  the deuterated forms of methanol.
\end{abstract}

\section{Introduction}
Methanol,CH$_3$OH, has been detected in several environments in the
interstellar medium, e.g. in dark clouds, pre-stellar cores, in the gas phase as
well as in ices.\cite{rat11} Like several other molecules, it is found to be heavily
deuterated. For example the ratio between CH$_3$OH and CD$_3$OH was found to
be merely 125 towards IRAS 16293-2422.\cite{par04} With a cosmic D/H abundance
ratio of $\sim 1.5\times 10^{-5}$,\cite{lin95} this corresponds to a deuterium
enrichment by a factor of over 13 orders of magnitude.
Moreover, deuteration is almost exclusively found on the C atom rather than on the OH
group.\cite{par04}

The reasons for that have been discussed in the past.\cite{nag05,nag06} Methanol
is mostly synthesized in the solid state on the surface of dust grains by
successive hydrogenation of carbon monoxide.\cite{hir94,wat03,wat04,gar07,fuc09,pir10,mor19} In that
process, the slowest reaction is the initial hydrogenation of CO to form HCO
or DCO. Here, formation of HCO is faster than the formation of DCO by factors between 10 and $>250$ at
low temperatures.\cite{hid07,wat08,and11} Formaldehyde, H$_2$CO, is
preferentially hydrogenated at the carbon end to form CH$_3$O.\cite{son17}
Also in this reaction, H-addition is much faster than D-addition,\cite{son17}
as are the majority of reactions that exhibit a barrier. Overall, we must
assume that methanol is initially produced even in a deuterium-depleted manner
with respect to the cosmic H/D ratio.

One way to explain the high degree of deuteration is the lower zero point
vibrational energy (ZPE) of, e.g., CD$_3$OH vs. CH$_3$OH. However, the
majority of deuterium is stored in the form of HD in the interstellar medium,
which has an even lower ZPE per hydrogen atom. Alternatively, H-D exchange in
solid methanol was suggested as an explanation for the selective
deuteration.\cite{nag05,nag06} Hydrogen atoms may abstract H or D from
methanol. This reaction will be the focus of this study. Here, the
H-abstraction is much faster than the D-abstraction, leading to a
D-enrichment.\cite{nag05,nag06,gou11a} The resulting CH$_2$OH can, again,
react with H or D in a barrier-less process to reform methanol. Since the
process is barrier-less, both H and D are expected to react almost equally
fast.  Both, abstraction and addition, are expected to happen on the surface
of dust grains, since the dust grain allows for the dissipation of the excess
energy from the radical-radical recombination, which is impossible in the gas
phase.

It was shown previously that abstraction from the CH$_3$ group of methanol,
resulting in CH$_2$OH, is much more likely than abstraction from the OH
group,\cite{len97,jod99,ker04,mea11} which would result in CH$_3$O. This
explains the selective deuteration. Moreover, CH$_2$OH is thermodynamically
more stable than CH$_3$O. A conversion from CH$_2$OH to
CH$_3$O is unlikely.\cite{wan12,rya12}

While the hydrogenation reactions of CO have been studied extensively,
studies of hydrogen abstraction from CH$_3$OH are rarer. Gas-phase
kinetic measurements for temperatures of 200~K and above are
available,\cite{mea74,bau05} as well as simulations.\cite{ker04,car08,mea11,san17,sha18}
In previous work,\cite{gou11a} the title reaction was investigated to
low temperature using instanton theory based on density functional
theory. To improve on the limits of that method, we now describe the
reaction based on coupled cluster theory. We provide rate constants
for several isotope patterns using instanton theory. Since on-the-fly
coupled cluster calculations would be computationally too demanding,
especially since we aim at multiple isotope patterns, we fitted the
potential energy surface by a neural network, as described
previously.\cite{coo18} We now explicitly study all H/D isotope
patterns on the CH$_3$ group as well as for the incoming H or D
atom. Moreover, we investigated the $^\text{12}$C/$^\text{13}$C and
$^\text{16}$O/$^\text{18}$O kinetic isotope effects (KIE). Since all
the reactions are expected to happen on the surface of dust grains, we
provide unimolecular rate constants, which describe the decay of a
pre-reactive complex between H and methanol on the surface in a
Langmuir--Hinshelwood process.\cite{mei17} \new{There are four main effects of
adsorption on a surface on such a reaction: (1) a higher concentration of
reactive species, (2) dissipation of excess energy of the reaction into the
surface, (3) influence of the surface environment on the reaction barrier
(catalytic effect), and (4) restriction of the rotation and translation of
the reactants on the surface. Effects 1 and 2 are handled implicitly by our
approach, since we calculate canonical rate constants, which are independent
of the concentrations and assume a thermalized ensemble. A catalytic effect of
the surface is expected to be small for the rather apolar surfaces on which
methanol is expected (dirty CO ice). Even for water ice surfaces, which are
much more polar, such effects were found to be small
\cite{lam16,son16,son17,lam17a,mei17}. Thus, we neglect catalytic surface
effects here. The rotation and translation of the methanol molecule are
restricted by keeping its rotational and translational partition functions
constant between the reactant and the instanton. Such an implicit surface
model, which allows the description of surface reactions by a gas-phase
structural model, was used successfully previously.\cite{mei17}}

The different deuteration patterns result in the following reactions, where
the incoming atom X can be H or D. The first three reactions involve H
abstraction, reactions \ref{rkn:4} to \ref{rkn:6} involve D abstraction.
\begin{reaction}
  \text{CH}_3\text{OH}+\text{X} \rightarrow
  \text{CH}_2\text{OH}+\text{XH}
  \label{rkn:1}
\end{reaction}
\begin{reaction}
  \text{CH}_2\text{D}\text{OH}+\text{X} \rightarrow
  \text{CHDOH}+\text{XH}.
  \label{rkn:2}
\end{reaction}
\begin{reaction}
  \text{CH}\text{D}_2\text{OH}+\text{X} \rightarrow
  \text{CD}_2\text{OH}+\text{XH}.
  \label{rkn:3}
\end{reaction}
\begin{reaction}
  \text{CD}\text{H}_2\text{OH}+\text{X} \rightarrow
  \text{CH}_2\text{OH}+\text{XD}.
  \label{rkn:4}
\end{reaction}
\begin{reaction}
  \text{CD}_2\text{H}\text{OH}+\text{X} \rightarrow
  \text{CDH}\text{OH}+\text{XD}.
  \label{rkn:5}
\end{reaction}
\begin{reaction}
  \text{CD}_3\text{OH}+\text{X} \rightarrow
  \text{CD}_2\text{OH}+\text{XD}.
  \label{rkn:6}
\end{reaction}
\begin{reaction}
  \ce{^{13}CH_3 OH + H} \rightarrow 
  \ce{^{13}CH_2 OH + H_2}
  \label{rkn:7}
\end{reaction}
\begin{reaction}
  \ce{CH_3 ^{18}OH + H} \rightarrow 
  \ce{CH_2 ^{18}OH + H_2}
  \label{rkn:8}
\end{reaction}

\section{Computational Details}

\subsection{Neural-Network Potential Energy Surface}

The neural-network potential energy surface (NN-PES) used in this paper has
been constructed as described in reference \citenum{coo18} where reaction \ref{rkn:1} was
studied to prove the applicability of a NN-PES for the accurate calculation of
reaction rate constants with the instanton method.  Therefore we will only
report the essential computational details on the training of the NN-PES here
and refer to Ref.~\citenum{coo18} for further details.

The training and test set used in this work were constructed on the basis of
the reference data used previously.\cite{coo18} The reference energies for the
training and test set structures were calculated with unrestricted explicitly
correlated coupled-cluster theory, where single and double excitations were
considered and triple excitations were treated perturbatively,
UCCSD(T)-F12/VTZ-F12, on a restricted Hartree-Fock basis. Compared to our
previous work,\cite{coo18} some slight adjustments have been made: First all
redundant structures, whose overall coordinates differed less than $10^{-2}$
Bohr were deleted. Subsequently several 
structures
from the minimum regions were added to the training set.  In total the
training set consists of 70 and the test set of 18 structures.  In order to
obtain a NN-PES that is suitable for the calculation of reaction rate
constants it is beneficial to directly include gradient and Hessian
information in the training process. Therefore, gradients and Hessians were
calculated for all reference structures by finite differences of the coupled
cluster energies employing a 4$^\text{th}$ order scheme.  All energy
calculations were performed using Molpro 2012.7\cite{wer12} via
ChemShell\cite{she03,met14} with an energy threshold of $10^{-10}$
Hartree. The gradients and Hessians were calculated in DL-FIND\cite{kae09} via
Chemshelll.\cite{she03,met14}

The coordinates were described by normal vibrational coordinates relative to
the transition state structure. A feed forward neural network with two hidden
layers was used and the network architecture employed was 15-50-50-1,
i.e. there are 15 nodes in the input layer (i.e. the 15 vibrational degrees of
freedom), 50 nodes in both hidden layers and one node in the output layer.
For training a batch training approach was chosen where the L-BFGS
algorithm\cite{liu89} was used to minimize the cost function.  

\subsection{Reaction Rate Constants}
Reaction rate constants including quantum mechanical tunneling were calculated
using instanton theory,\cite{lan67,mil75,col77,cal77,ric09,alt11,kae14,ric16}
which is based on a semi-classical approximation of the Feynman path integral
formalism. Instanton theory in its standard formulation is only applicable for
temperatures below the crossover temperature $T_\text{c}=\frac{\hbar
  \omega_\text{TS}}{2 \pi k_\text{B}}$, where $\hbar$ is the reduced Planck
constant, $\omega_\text{TS}$ is the absolute value of the imaginary frequency
at the transition structure and $k_\text{B}$ is Boltzmann's constant. Thereby
the crossover temperature $T_\text{c}$ is mass dependent since
$\omega_\text{TS}$ is mass dependent.  The instanton is at a given temperature
the tunneling path with the highest statistical weight. The instanton was
located using a modified Newton--Raphson method\cite{rom11,rom11b} in
DL-FIND.\cite{kae09} Close to $T_\text{c}$, instanton theory is known to
overestimate the rate constant.\cite{kae14} Thus, a
correction\cite{kry11,mcc17a} was applied. When taking the rotational symmetry
factors\cite{fer07a} into account, the three hydrogen atoms at the CH$_3$
group were considered indistinguishable, unless they were different isotopes.

To model Langmuir--Hinshelwood processes, unimolecular rate constants were
calculated for all isotope patterns considered \new{using the implicit surface model}.\cite{mei17} For completeness,
the bimolecular gas-phase rate constants for \ref{rkn:1} with an incoming H
atom were also calculated. For these calculations, the rate constants were
obtained from a microcanonical formulation of instanton
theory,\cite{mil75,ric16a,mcc17,loh18} which is the appropriate model for the
low-pressure limit, in which thermal equilibration in the pre-reactive complex
is excluded. The microcanonical rate constants were obtained from solving the
stability matrix differential equation.\cite{mcc17} In this work the
instantons were discretized to 200 images and the convergence criterion for
the instanton was set to the gradient of the Euclidean action $S_\text{E}$
with respect to the mass-weighted coordinates being less than $5.0\cdot
10^{-11}$ atomic units. 
\section{Results}

\begin{figure}[h]
  \centering
  \includegraphics[width=8cm]{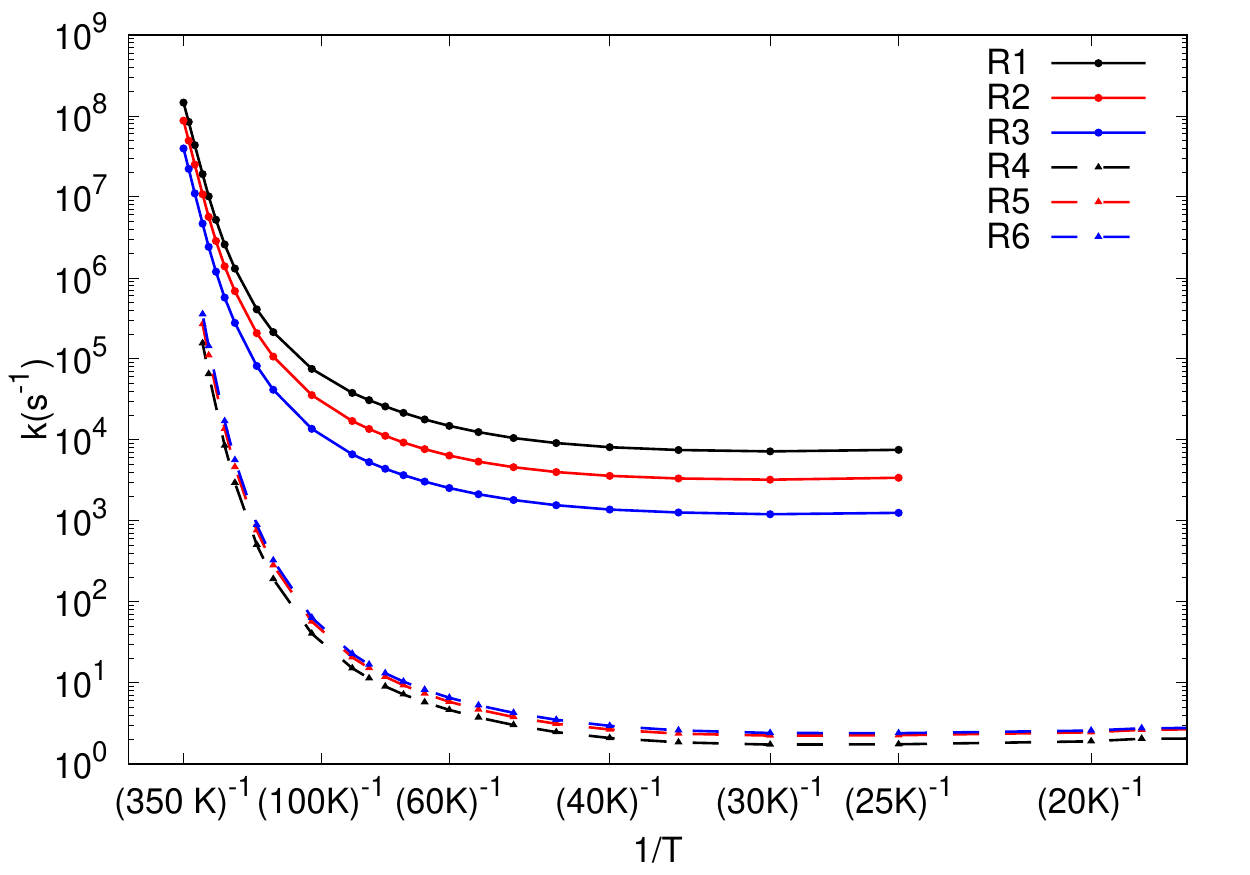}
  \caption{Rate constants for the reactions \ref{rkn:1}--\ref{rkn:6} for an incoming H atom. 
  }
  \label{fig:inc_H}
\end{figure}

\begin{figure}[h]
  \centering
  \includegraphics[width=8cm]{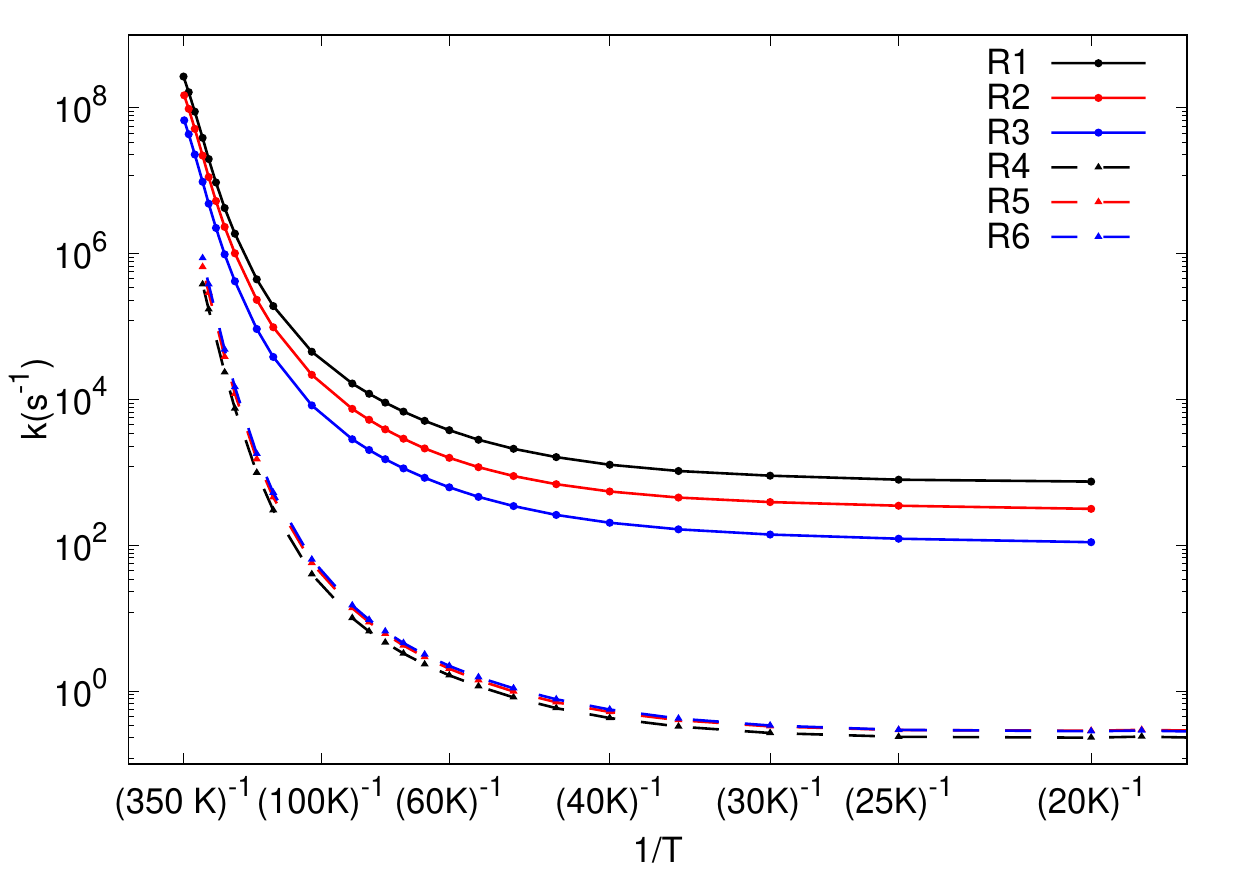}
  \caption{Rate constants for the reactions \ref{rkn:1}--\ref{rkn:6} for an incoming D atom.
  \label{fig:inc_D}}
\end{figure}

Temperature-dependent rate constants for the 12 H/D combinations of reactions
\ref{rkn:1}--\ref{rkn:6} were calculated. Graphs are shown in
Figures \ref{fig:inc_H} and \ref{fig:inc_D}, numbers for some temperatures are
given in Tables \ref{tbl:inc_H} and \ref{tbl:inc_D}. A more extensive list of
rate constants is given in the Supporting Information.
Overall, one can see that
the unimolecular rate constants are almost temperature-independent below
40~K. Below that temperature, all of the reactions are dominated by tunneling
from the ground state of the reactant state complex. Primary KIEs are
substantial: the replacement of the abstracted H by D decreases the rate
constant by a factor of about 3000 to 4000 at 30~K,
depending on the H/D pattern of the other atoms. This ratio has a direct
consequence for the deuterium fractionation of methanol, as shall be discussed
below. Changing the abstracting atom from H to D also decreases the rate
constant, but to a much lesser degree, by factors of about 6 to 9 at
30~K. \new{This is caused by two opposing contributions: while tunneling
  decreases the rate constant for abstraction by D, the vibrational zero point
  energy increases it, see Table~\ref{tbl:inc_D}. The increase in mass of the
  abstracting atom hardly changes the ZPE of the reactant, in which that atom
  is bound only weakly, while it reduces the ZPE of the transition state.}
 
Secondary KIEs are smaller. The rate of abstracting a
H atom from CH$_3$OH by a H atom is about 6 times higher than from
CHD$_2$OH at 30~K. Of that, a factor of 3 originates merely from the
rotational symmetry factor, which essentially captures the fact that there are
three H atoms to abstract in CH$_3$OH, while there is only one in
CHD$_2$OH. Only the remaining factor of 1.99 is caused by the
different masses.

\begin{table}[h]
\small
  \caption{Data for the reactions \ref{rkn:1} to \ref{rkn:8} for an incoming H atom. 
  $E_\text{uni,act}$ refers to the unimolecular activation energy including ZPE, 
  $T_\text{c}$ is the crossover temperature. The KIE is given with respect to H-\ref{rkn:1}, 
  values in parentheses refer to powers of 10.
}
  \label{tbl:inc_H}
  \begin{tabular*}{0.75\textwidth}{lc@{ }cr@{}lr@{}lc}
    \hline
    Reactions & $E_\text{uni,act}$ & $T_\text{c}$ & \multicolumn{4}{c}{KIE
      w.r.t. H-R1} & $k$ at 30~K\\
    &(kJ/mol)& (K) & 105~K & & 30~K & &(s$^{-1}$)\\
    \hline
    \ref{rkn:1}: $\text{CH}_3\text{OH}+\text{H}$         & 33.1 & 357 &     &    &     &     & 7.22(3) \\
    \ref{rkn:2}: $\text{CH}_2\text{D}\text{OH}+\text{H}$ & 33.2 & 356 &    2&.11 &    2&.24  & 3.23(3) \\
    \ref{rkn:3}: $\text{CH}\text{D}_2\text{OH}+\text{H}$ & 33.4 & 355 &    5&.50 &    5&.96  & 1.21(3) \\
    \ref{rkn:4}: $\text{CD}\text{H}_2\text{OH}+\text{H}$ & 37.9 & 269 & 1850&    & 4170&     & 1.73(0) \\
    \ref{rkn:5}: $\text{CD}_2\text{H}\text{OH}+\text{H}$ & 38.1 & 269 & 1310&    & 3230&     & 2.23(0) \\
    \ref{rkn:6}: $\text{CD}_2\text{H}\text{OH}+\text{H}$ & 38.3 & 268 & 1180&    & 3000&     & 2.41(0) \\
    \ref{rkn:7}: \ce{^{13}CH_3 OH + H}                   & 33.1 & 356 &    1&.06 &    1&.08  & 6.69(3) \\
    \ref{rkn:8}: \ce{CH_3 ^{18}OH + H}                   & 33.1 & 357 &    1&.00 &    0&.998 & 7.24(3) \\
    \hline
  \end{tabular*}
\end{table}

\begin{table}[h]
\small
  \caption{Data for the reactions \ref{rkn:1} to \ref{rkn:8} for an incoming D atom. 
  $E_\text{uni,act}$ refers to the unimolecular activation energy including ZPE, 
  $T_\text{c}$ is the crossover temperature. The KIE is given with respect to D-\ref{rkn:1}, 
  values in parentheses refer to powers of 10.
  \label{tbl:inc_D}
  }
    \begin{tabular*}{0.75\textwidth}{l@{ }c@{ }cr@{}lr@{}lc}
    \hline
    Reactions & $E_\text{uni,act}$ & $T_\text{c}$ &  \multicolumn{4}{c}{KIE
      w.r.t. D-R1} & $k$ at 30~K\\
    & (kJ/mol) & (K) & 105~K & & 30~K & &(s$^{-1}$)\\
    \hline
    \ref{rkn:1}: $\text{CH}_3\text{OH}+\text{D}$         & 30.3 & 353 &     &    &     &    & 8.97(2)  \\
    \ref{rkn:2}: $\text{CH}_2\text{D}\text{OH}+\text{D}$ & 30.4 & 351 &    2&.08 &    2&.30 & 3.90(2)  \\
    \ref{rkn:3}: $\text{CH}\text{D}_2\text{OH}+\text{D}$ & 30.5 & 350 &    5&.44 &    6&.41 & 1.40(2)  \\
    \ref{rkn:4}: $\text{CD}\text{H}_2\text{OH}+\text{D}$ & 35.0 & 265 & 1130&    & 3380&    & 2.66($-$1) \\
    \ref{rkn:5}: $\text{CD}_2\text{H}\text{OH}+\text{D}$ & 35.2 & 264 &  791&    & 2740&    & 3.27($-$1) \\
    \ref{rkn:6}: $\text{CD}_2\text{H}\text{OH}+\text{D}$ & 35.4 & 263 &  709&    & 2660&    & 3.37($-$1) \\
    \hline
  \end{tabular*}
\end{table}

\begin{figure}[h]
  \centering
  \includegraphics[width=8cm]{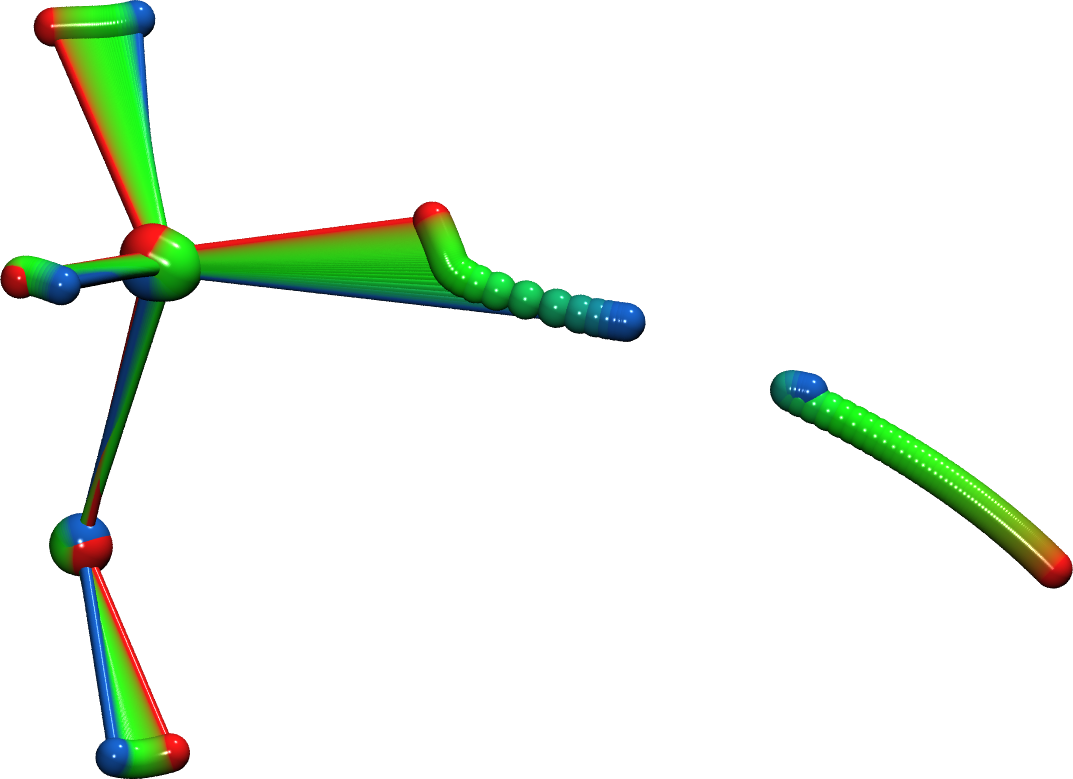}
  \caption{Instanton path for the reaction \ref{rkn:1}-H at 30~K. The red atom
    positions refer to the reactant state CH$_3$OH + H, the blue positions
    refer to the turning point of the instanton path at the product side,
    CH$_2$OH + H$_2$. The OH group is found towards the bottom of the image.}
  \label{fig:instpath}
\end{figure}

\begin{figure}[h]
  \centering
  \includegraphics[width=8cm]{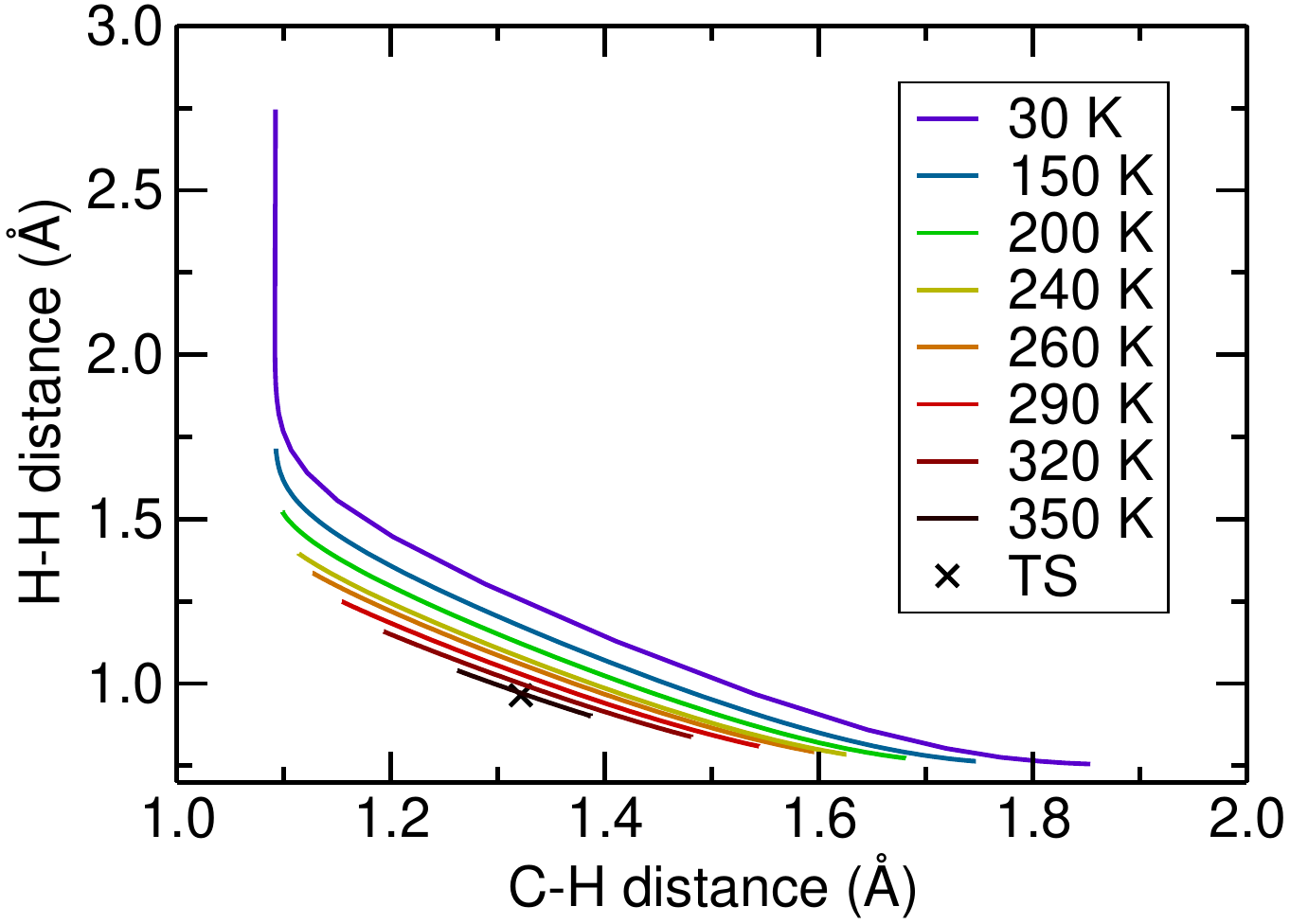}
  \caption{\new{Instanton paths for  \ref{rkn:1}-H at different
      temperatures. The changes in the C--H and H--H distances of the
      abstraction are displayed.}
    \label{fig:instpath:temp}}
\end{figure}

All atoms in the reaction contribute to the tunneling path, as can be seen
from the instanton path for reaction \ref{rkn:1} with an incoming H atom,
illustrated in Figure \ref{fig:instpath} for $T=30$~K. The instanton path is a
closed Feynman path, which re-traces itself between two turning points. At low
temperature, one turning point gets close to the reactant minimum. In Figure
\ref{fig:instpath}, this is given by the red geometry. The other turning
point (blue) has the same energy as the first one, but is within the product
valley. \new{The changes of the C--H and H--H distances along the
  instanton are displayed in Figure~\ref{fig:instpath:temp}. At high
  temperature, the instanton is short and close to the classical
  transition structure. Lowering the temperature leads to 
  longer instnaton paths.}
The path length of the incoming H atom is 1.31~\AA, while the path
length of the abstracted H atom is 0.92~\AA. The secondary hydrogen atoms have
much shorter path lengths of 0.18, 0.22 and 0.32~\AA. Replacing an atom with a
heavier isotope results in an effective shortening of its contribution to the
instanton path. While the path of the incoming hydrogen atom can be shortened
without raising the energy much (the potential energy surface is rather flat
in that area), changes to the abstracted atom's path has a huge influence on
the energy. This is the reason why the KIE with respect to the incoming atom
is rather small (one order of magnitude), while the KIE with respect to
changing the abstracted atom is huge (3.5 orders of magnitude). 

\begin{figure}[h]
  \centering
  \includegraphics[width=8cm,clip]{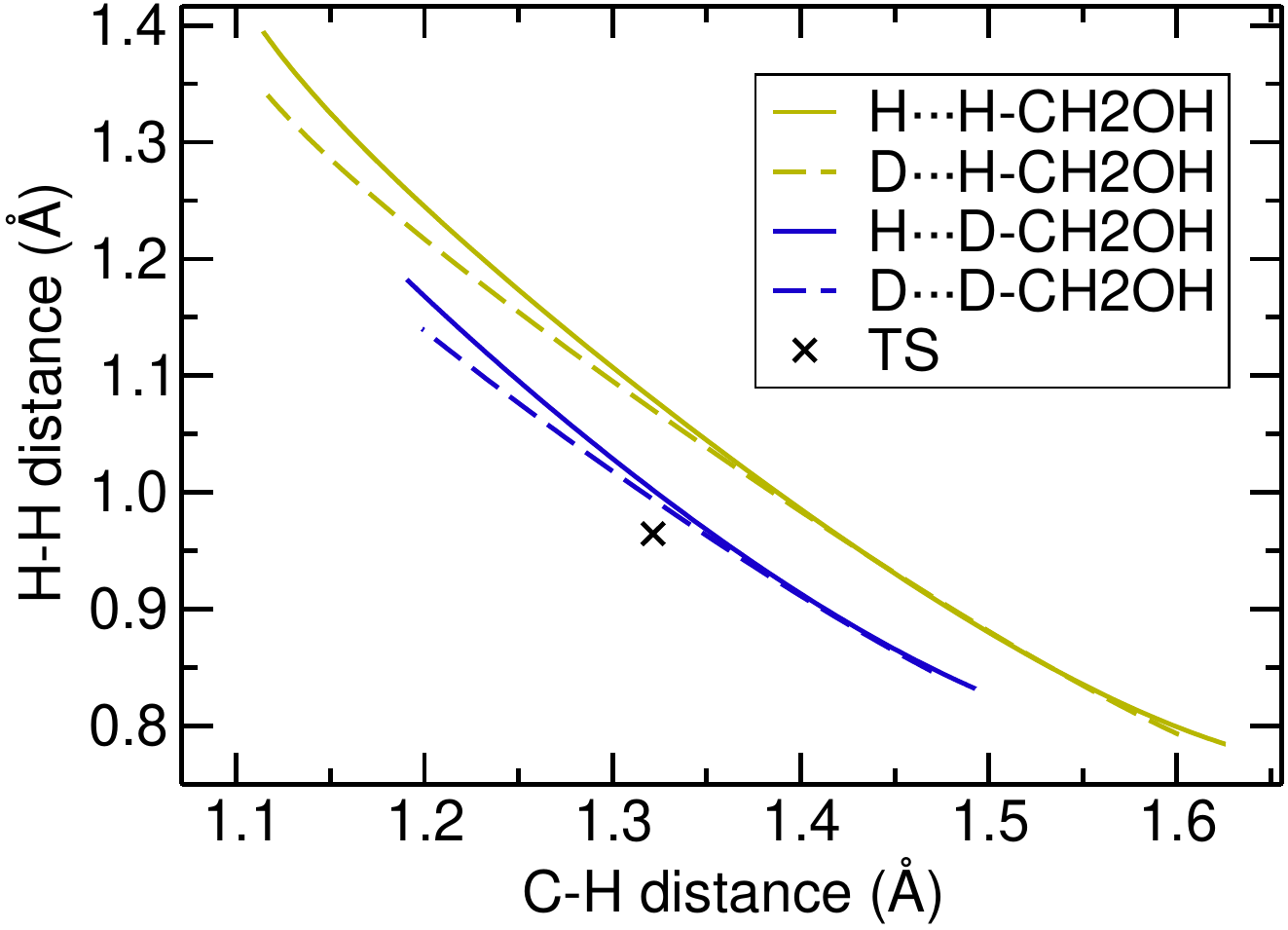}
  \caption{\new{Instanton paths for different mass combinations
      at 240~K.}
    \label{fig:path:mass}}
\end{figure}

\new{The influence of deuterium substitution on the C--H and H--H distances along the
  instanton are displayed in Figure~\ref{fig:path:mass}. An increase in
  the mass leads to a shortening of the path. The H--H distance is
  noticeably reduced by an increase of the mass in the abstracting
  hydrogen atom, especially in the reactant state region (upper left),
  where the potential energy is flat.}

\begin{figure}[h]
  \centering
  \includegraphics[width=8cm]{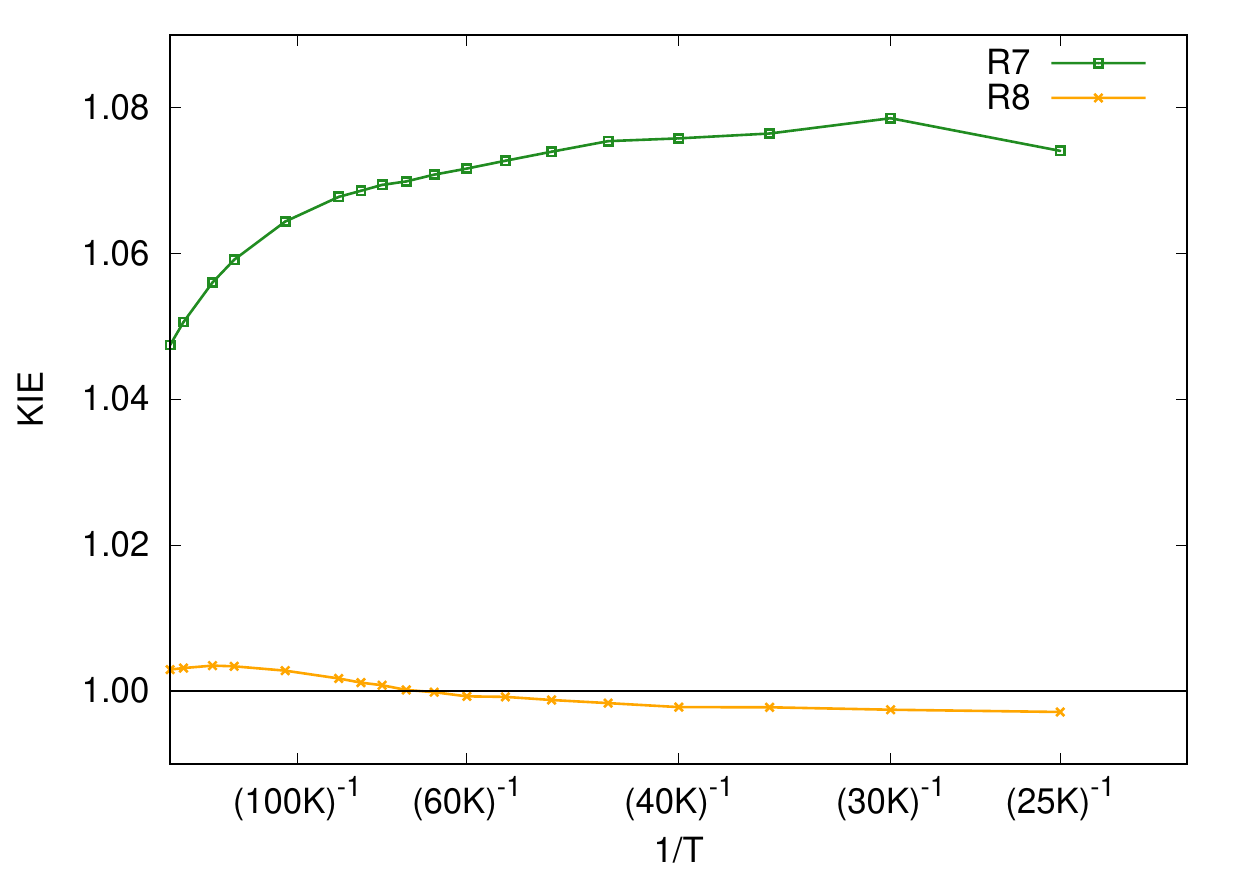}
  \caption{Heavy-atom KIEs for the title reaction, $^\text{12}$C/$^\text{13}$C
    in green and $^\text{16}$O/$^\text{18}$O in yellow. 
  \label{fig:kieCO}}
\end{figure}

Heavy-atom KIEs were also calculated. They are, naturally, much smaller than
the H/D KIEs. Replacing $^\text{12}$C by $^\text{13}$C leads to a KIE of
1.0786 at 30~K. Somewhat smaller values are found at higher temperature, see Figure
\ref{fig:kieCO}. The difference in mass between the carbon isotopes is
small, but the atom is involved significantly in the tunneling process. During
the reaction, the C atom moves towards the abstracted H atom, afterwards it
moves back. Its path length is still 0.15~\AA{} for \ref{rkn:1}-H at
30~K. Much smaller effects are seen for an $^\text{16}$O/$^\text{18}$O
replacement. At 30~K, we even obtain an inverse KIE of 0.9974. The oxygen atom
hardly moves during the reaction, with an instanton path length of only
0.03~\AA.

\begin{figure}[h]
  \centering
  \includegraphics[width=8cm]{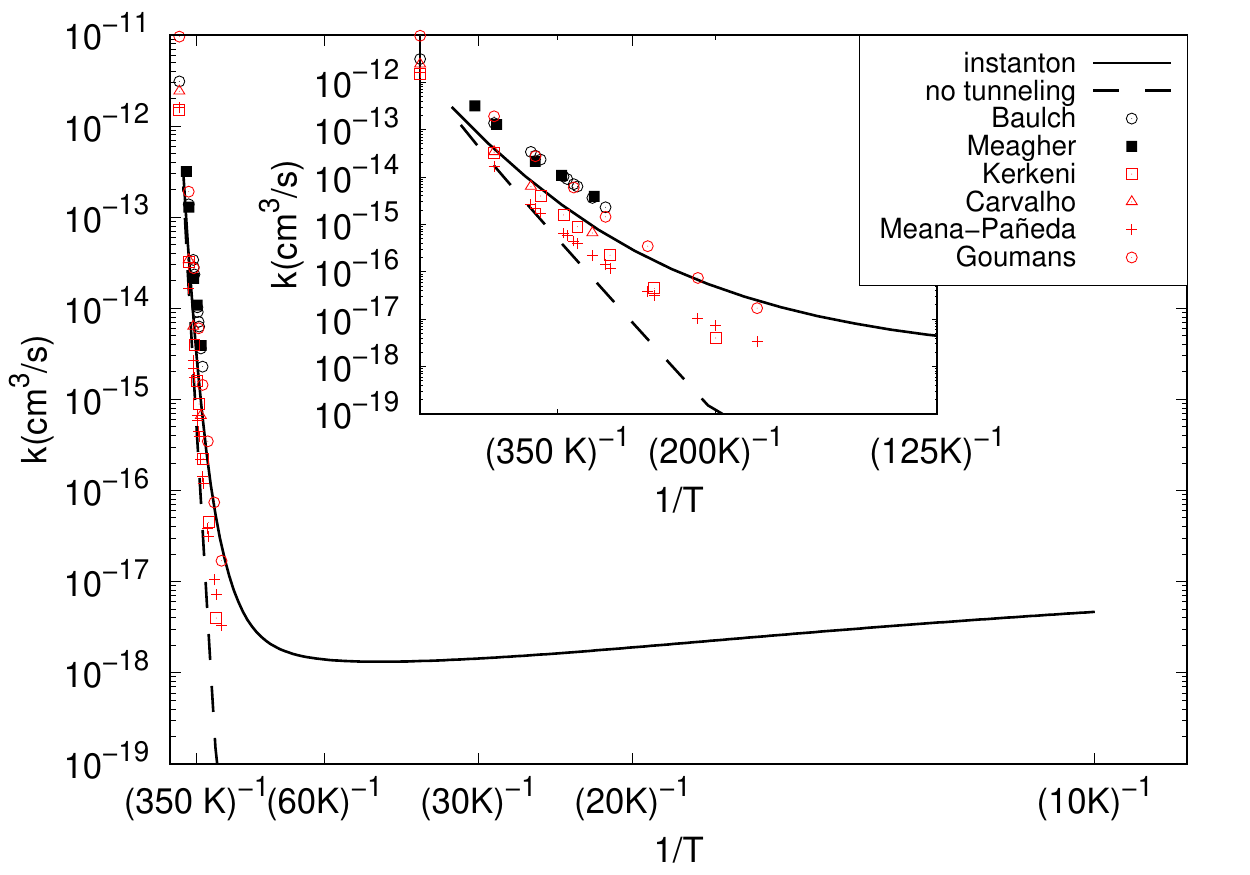}
  \caption{Bimolecular rate constants for \ref{rkn:1}-H obtained from a
    microcanonical formulation (solid lines). Literature data are also
    provided. 
    \label{fig:micro2}}
\end{figure}

The unimolecular rate constants presented so far refer to the thermalized
surface process, i.e. a Langmuir--Hinshelwood mechanism on the surface. The
abstraction reaction may, however, also happen in the gas phase. Reactions
with a pre-reactive minimum lead to technical difficulties in canonical
instanton theory, as that would assume that the pre-reactive complex is
thermalized.\cite{mcc17} At low pressure, however, such a thermalization is unlikely. The correct treatment
of low-pressure bimolecular processes is to calculate microcanonical rate
constants (cumulative reaction probabilities) and use those to calculate
thermal rate constants using a thermal ensemble of the separated
reactants. This was done to obtain the rate constants shown in
Figure \ref{fig:micro2}. The rate constants steeply decrease until tunneling sets
in at about the crossover temperature of 357~K. At very low temperature, the
rate constant slightly increases again, a fact that can be seen for many
bimolecular cases. It is caused by a delicate balance
between the lost rotational and translational degrees of freedom and the
additional vibrational degrees of freedom when the two reactants form one
transition state.
At high temperature our rate constants compare fairly well with the experimental data by Meagher et
al.\cite{mea74} and the experiment-based recommendations by Baulch et
al.\cite{bau05} Also at high temperature, we can compare to the
simulation results
by  Kerkeni and Clary obtained using quantum dynamics,\cite{ker04} the
VTST/ZCT values by Carvalho et al.,\cite{car08} or the expression fitted
to VTST/$\mu$OMT data by Meana-Pa{\~n}eda et al.\cite{mea11} and the
DFT-based instanton data.\cite{gou11a} The
comparision is generally quite goood, see Figure \ref{fig:micro2}. To the
best of our knowledge, no data are available in the literature below 180~K.

\section{Discussion}

\begin{figure}[h]
  \centering
  \includegraphics[width=8cm]{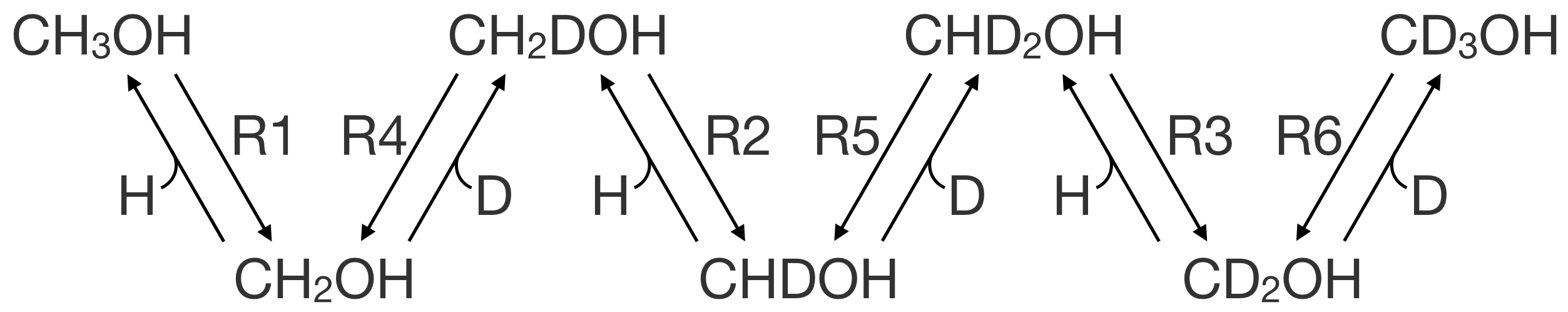}
  \caption{Scheme of the kinetic model to explain the contribution of
    the title reaction to deuteration of methanol. Isotopologues of
    methanol are shown in the top row, isotopologues of the CH$_2$OH
    radical in the bottom row.
    \label{fig:kinmodel}}
\end{figure}

In order to gauge the influence of the title reaction on the deuterium
fractionation in methanol, we build a very small and simplistic chemical
network, only consisting of hydrogen (H or D) abstraction from methanol and
recombination of the resulting radical with hydrogen atoms. This is solved in
a steady-state model. A similar attempt was made previously.\cite{gou11a}
Here, we use the new rate constants for reactions \ref{rkn:1}--\ref{rkn:6},
which are based on more accurate electronic structure calculations and which
take secondary kinetic isotope effects into account. We, moreover take surface
diffusion into account, albeit in a rather simple manner.  The chemical
network studied is shown schematically in Figure \ref{fig:kinmodel}.

For the reaction of the closed-shell isotopologues of CH$_3$OH with H we
assume that the diffusion (hopping) rate constant of H atoms on the surface is
much higher than the rate constant of the chemical process, i.e. that the
process is reaction-limited. This assumption seems valid, since typical H
hopping rate constants are $10^6$ to $10^{11}$~s$^{-1}$,\cite{sen17,asg17}
while the rate constants we obtain for the chemical step are below
$10^4$~s$^{-1}$ at low temperature. The reaction-limited Langmuir--Hinshelwood
rate for reaction \ref{rkn:1} is\cite{mei17}
\begin{equation} 
  R_\text{LH,reaction-limited}=k_1 \frac{n(\text{H})
    n(\text{CH}_3\text{OH})}{n(\text{sites})}. \label{eq:1}
\end{equation}
Here $k_1$ is the rate constant of \ref{rkn:1}, $n(\text{H})$ is the surface
concentration of hydrogen atoms, $n(\text{CH}_3\text{OH})$ is the
surface concentration of methanol, and $n(\text{sites})$ is the surface
concentration of binding sites. Surface concentrations may be numbers of
sites/species, number densities, or real concentrations. We neglect
abstraction by D atoms, since they are much rarer than H atoms. Equations
similar to \eqref{eq:1} can be set up for \ref{rkn:2} to \ref{rkn:6}.

The recombinations of radicals with H or D atoms are barrier-less. Thus, they will
be diffusion-limited. Their rate constants are\cite{mei17}
\begin{equation} 
R_\text{LH,diffusion-limited}=[k_\text{diff}(\text{H})
+k_\text{diff}(\text{CH}_2\text{OH})]   \frac{n(\text{H}) n(\text{CH}_2\text{OH})}{n(\text{sites})}
\end{equation}
for reactions with H and 
\begin{equation} 
R_\text{LH,diffusion-limited}=[k_\text{diff}(\text{D})
+k_\text{diff}(\text{CH}_2\text{OH})]   \frac{n(\text{D}) n(\text{CH}_2\text{OH})}{n(\text{sites})}
\end{equation}
for reactions with D.  Here, $k_\text{diff}$ are the hopping rate constants on
the surface. At temperatures of 30~K or below, the diffusion of CH$_2$OH or
its isotopologues is much slower than the diffusion of H or D, thus
$k_\text{diff}(\text{CH}_2\text{OH})$ can be neglected.

\begin{figure}[h]
  \includegraphics[width=8cm]{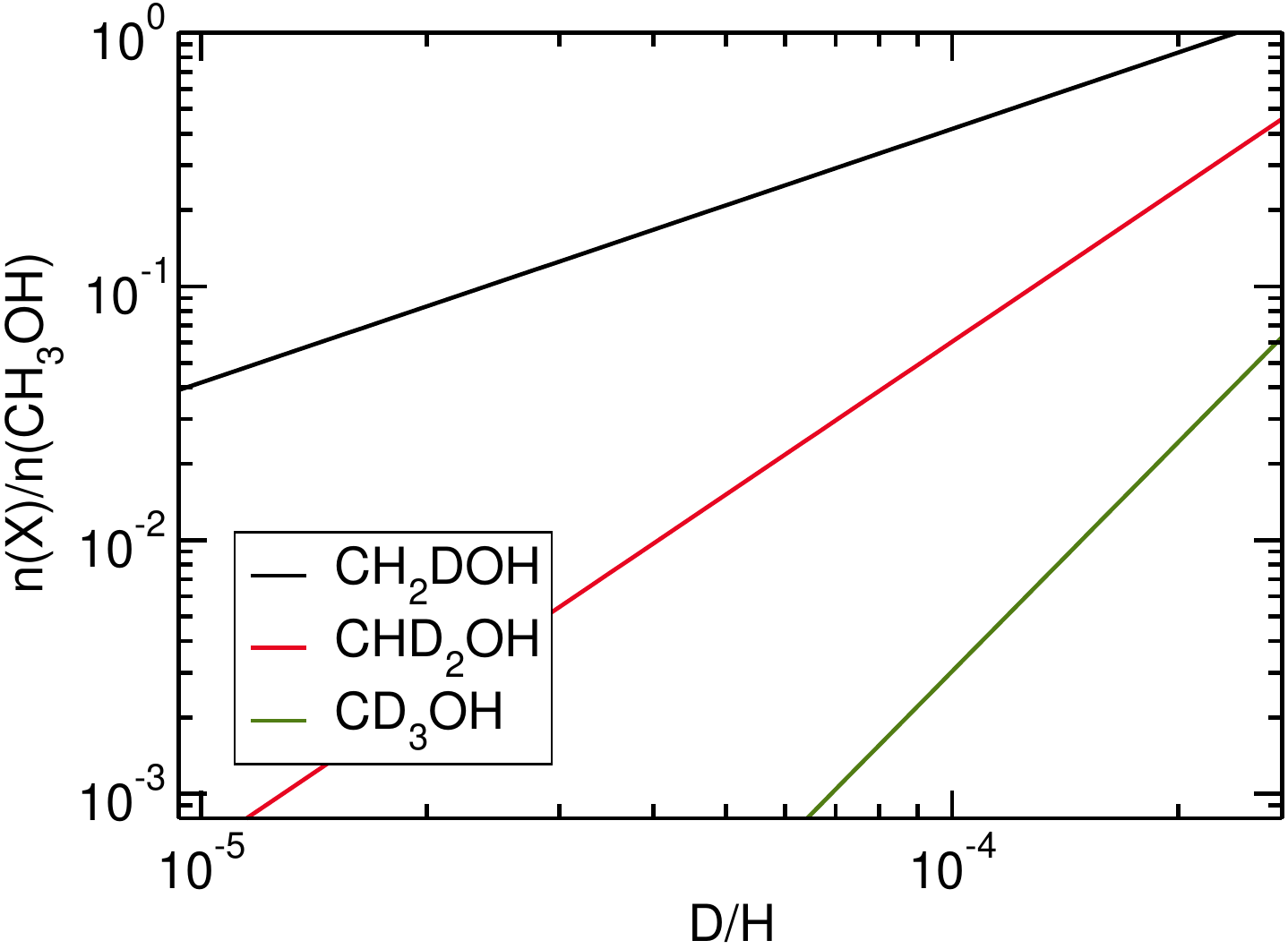}
  \caption{Calculated abundances of CH$_2$DOH, CHD$_2$OH, and
    CD$_3$OH relative to CH$_3$OH (solid lines).
    \label{fig:deutfrac}}
\end{figure}

Using these expressions, a set of differential equations can be constructed for
the kinetics of the model illustrated in Figure \ref{fig:kinmodel}:
\begin{equation}
\frac{d n(\text{CH}_3\text{OH})}{dt}=
-k_1 \frac{n(\text{H})n(\text{CH}_3\text{OH})}{n(\text{sites})} 
+ k_\text{diff}(\text{H}) \frac{n(\text{H}) n(\text{CH}_2\text{OH})}{n(\text{sites})}
\end{equation}
\begin{align}
\frac{d n(\text{CH}_2\text{OH})}{dt}=
k_1 \frac{n(\text{H})n(\text{CH}_3\text{OH})}{n(\text{sites})} 
- k_\text{diff}(\text{H}) \frac{n(\text{H}) n(\text{CH}_2\text{OH})}{n(\text{sites})}
\nonumber\\
+k_4 \frac{n(\text{H})n(\text{CH}_2\text{DOH})}{n(\text{sites})} 
- k_\text{diff}(\text{D}) \frac{n(\text{D}) n(\text{CH}_2\text{OH})}{n(\text{sites})}
\label{eq:4}
\end{align}
which can be continued in a similar manner for the other species. With all
time-derivatives vanishing in the steady-state assumption, the model
simplifies significantly, resulting in:
\begin{eqnarray}
n(\text{CH}_2\text{DOH})&=&\frac{k_1}{k_4}\frac D H n(\text{CH}_3\text{OH}) \label{eq:CH2DOH}\\
n(\text{CHD}_2\text{OH})&=&\frac{k_2}{k_5}\frac D H n(\text{CH}_2\text{DOH} \label{eq:CHD2OH}) \\
n(\text{CD}_3\text{OH})&=&\frac{k_3}{k_6} \frac D H n(\text{CHD}_2\text{OH} \label{eq:CD3OH})
\end{eqnarray}
where the short-hand notation $\frac D
H=\frac{k_\text{diff}(\text{D})n(\text{D})}
{k_\text{diff}(\text{H})n(\text{H})}$ is the ratio of the availabilities of H
and D. The ratio of diffusion constants was simulated on crystalline water and
amorphous water\cite{sen17,asg17} and measured on different
surfaces.\cite{ham12,ham13} It can be approximated to about 0.1.  The
concentrations of the isotopologues of methanol are independent of the
diffusion constants and of the intermediate concentrations of the radical
species. The concentrations of the radical species depend on the diffusion
constants. Using our rate constants at 30~K from Table \ref{tbl:inc_H}, we can
relate the deuterium fractionations of methanol to the availability of H vs. D
on the surface, see Figure \ref{fig:deutfrac}.

Observations toward IRAS 16293-2422 lead to abundances of CH$_2$DOH,
CHD$_2$OH, and CD$_3$OH relative to CH$_3$OH of $30 \pm 20$\%, $6 \pm 5$\%,
and $0.8 \pm 0.6$\%, respectively.\cite{par04}.
More recent observations towards the prototypical pre-stellar core L1544 lead to much smaller, but
spatially variable values of $8\pm 2$\% for
CH$_2$DOH/CH$_3$OH.\cite{biz14,cha19} All observations would correspond to
somewhat high D/H ratios in our model. 

Obviously, this is a very simplistic treatment. \new{The steady-state
  assumption was made ad-hoc to simplify the analysis.} A full kinetic model would
take many more reactions into account, primarily hydrogen (H and D)
abstraction by the radical-radical recombination. \new{Several other effects
  were neglected, which may have an influence on the rate constants. Any
  influence of the surface that changes the reaction barrier was neglected in
  the implicit surface model.\cite{mei17} All vibrations perpendicular to the instanton
  path were treated harmonically. Anharmonicity effects on the vibrational zero point
  energy were recently shown to influence the low-temperature rate constants
  of the reaction $\text{CH}_3\text{OH}\,+\,\text{OH}$.\cite{gao18}}

Nevertheless, this chemical model represents correct trends in the abundances of deuterated species of 
methanol. These can be directly related to the strong primary kinetic isotope effects found for the
exchange of the abstracted H atom with a D atom.
Solving equation~\ref{eq:CH2DOH} for $n(\text{CH}_2\text{DOH})/n(\text{CH}_3\text{OH})$ yields:
\begin{equation}
 \frac{n(\text{CH}_2\text{DOH})}{n(\text{CH}_3\text{OH})} = \frac{k_1}{k_4} \frac D H, 
\end{equation}
where $k_1/k_4$ is the kinetic isotope effect for \ref{rkn:4} with respect to H-\ref{rkn:1} at 30 K.
This reformulation shows clearly that the KIE defines the slope
of the corresponding straight line shown Figure \ref{fig:deutfrac}.
Therefore it becomes obvious that the reason for the concentration of $\text{CH}_2\text{DOH}$ relative 
to the concentration of CH$_3$OH being
three orders of magnitude higher than the concentration one would expect for a statistical exchange of H with D 
according to a uniform distribution for a given D/H ratio is the strong kinetic isotope effect of 4170 at 
30~K. Reformulations of equations \ref{eq:CHD2OH} and \ref{eq:CD3OH} lead to the same conclusion that  
the KIE for \ref{rkn:5} or respectively \ref{rkn:6} with respect to H-\ref{rkn:1} at 30 K dominates the
slope of the corresponding straight line in Figure \ref{fig:deutfrac} and thus give a good qualitative
explanation for the unintuitively high concentrations of deuterated methanol species, that were observed
in the interstellar medium.


\section{Conclusions}
We have studied the kinetics of $\text{H}+\text{CH}_3\text{OH} \rightarrow \text{H}_2 + \text{CH}_2\text{OH}$
and all H/D isotope patterns on the CH$_3$ group.
This study provides unimolecular reaction rate constants relevant for surface reactions, 
that model Langmuir--Hinshelwood processes, down to temperatures as low as 25~K.
Thereby the rate constants were calculated on a neural network potential energy surface that was 
fitted to UCCSD(T)-F12/VTZ-F12 data.
For the unimolecular rate constants also kinetic isotope effects are given for all studied isotope patterns 
at 105~K and 30~K.
Our study shows that primary KIEs are substantial for all isotope patterns. Replacing the abstracted H by D 
decreases the rate constants by a factor of 3000 to 4000 at 30~K. It was further shown that exchanging the 
abstracting H atom with a D atom  leads to a decrease in the rate constant by a factor of 6--9 at 30~K.
In this study also heavy-atom KIEs  were computed. The KIE for replacing 
$^\text{12}$C by $^\text{13}$C is 1.0786 at 30~K whereas the KIE for exchanging $^\text{16}$O with $^\text{18}$O  is vanishingly 
small. On top of that also secondary KIEs are given for all studied isotope patterns.
Further, bimolecular rate constants derived from a microcanonic formulation, that are relevant for the description of reactions 
in the gas-phase in the low pressure limit, are given down to unsurpassedly low temperatures (10~K).
With the help of a simplistic kinetic model it was found that the strong primary KIEs for replacing
the abstracted H atom by D are a good qualitative explanation for the unexpectedly 
high concentrations of deuterated methanol species that were found experimentally in the interstellar medium.

\begin{acknowledgement}
This work was financially supported by the European Union's Horizon 2020
research and innovation programme (grant agreement No. 646717, TUNNELCHEM) and
the German Research Foundation (DFG) via the grant SFB 716/C.6. Computational
resources were provided by the state of Baden-W\"urttemberg through bwHPC and
the German Research Foundation (DFG) through grant no INST 40/467-1 FUGG.
\end{acknowledgement}

\begin{suppinfo}

The following files are available free of charge.
\begin{itemize}
  \item supporting\_info.pdf: This file contains for all reactions discussed a detailed list 
   of bimolecular as well as unimolecular reaction rate constants for the whole temperature range
   covered in this study. 
\end{itemize}

\end{suppinfo}

\bibliography{new}

\providecommand{\latin}[1]{#1}
\providecommand*\mcitethebibliography{\thebibliography}
\csname @ifundefined\endcsname{endmcitethebibliography}
  {\let\endmcitethebibliography\endthebibliography}{}
\begin{mcitethebibliography}{62}
\providecommand*\natexlab[1]{#1}
\providecommand*\mciteSetBstSublistMode[1]{}
\providecommand*\mciteSetBstMaxWidthForm[2]{}
\providecommand*\mciteBstWouldAddEndPuncttrue
  {\def\EndOfBibitem{\unskip.}}
\providecommand*\mciteBstWouldAddEndPunctfalse
  {\let\EndOfBibitem\relax}
\providecommand*\mciteSetBstMidEndSepPunct[3]{}
\providecommand*\mciteSetBstSublistLabelBeginEnd[3]{}
\providecommand*\EndOfBibitem{}
\mciteSetBstSublistMode{f}
\mciteSetBstMaxWidthForm{subitem}{(\alph{mcitesubitemcount})}
\mciteSetBstSublistLabelBeginEnd
  {\mcitemaxwidthsubitemform\space}
  {\relax}
  {\relax}

\bibitem[Ratajczak \latin{et~al.}(2011)Ratajczak, Taquet, Kahane, Ceccarelli,
  Faure, and Quirico]{rat11}
Ratajczak,~A.; Taquet,~V.; Kahane,~C.; Ceccarelli,~C.; Faure,~A.; Quirico,~E.
  The Puzzling Deuteration of Methanol in Low- to High-Mass Protostars.
  \emph{Astron. Astrophys.} \textbf{2011}, \emph{528}, L13\relax
\mciteBstWouldAddEndPuncttrue
\mciteSetBstMidEndSepPunct{\mcitedefaultmidpunct}
{\mcitedefaultendpunct}{\mcitedefaultseppunct}\relax
\EndOfBibitem
\bibitem[Parise \latin{et~al.}(2004)Parise, Castets, Herbst, Caux, Ceccarelli,
  Mukhopadhyay, and Tielens]{par04}
Parise,~B.; Castets,~A.; Herbst,~E.; Caux,~E.; Ceccarelli,~C.;
  Mukhopadhyay,~I.; Tielens,~A. First Detection of Triply-Deuterated Methanol.
  \emph{Astron. Astrophys.} \textbf{2004}, \emph{416}, 159\relax
\mciteBstWouldAddEndPuncttrue
\mciteSetBstMidEndSepPunct{\mcitedefaultmidpunct}
{\mcitedefaultendpunct}{\mcitedefaultseppunct}\relax
\EndOfBibitem
\bibitem[Linsky \latin{et~al.}(1995)Linsky, Diplas, Wood, Brown, Ayres, and
  Savage]{lin95}
Linsky,~J.~L.; Diplas,~A.; Wood,~B.~E.; Brown,~A.; Ayres,~T.~R.; Savage,~B.~D.
  Deuterium and the Local Interstellar Medium Properties for the Procyon and
  Capella Lines of Sight. \emph{Astrophys. J.} \textbf{1995}, \emph{451},
  335\relax
\mciteBstWouldAddEndPuncttrue
\mciteSetBstMidEndSepPunct{\mcitedefaultmidpunct}
{\mcitedefaultendpunct}{\mcitedefaultseppunct}\relax
\EndOfBibitem
\bibitem[Nagaoka \latin{et~al.}(2005)Nagaoka, Watanabe, and Kouchi]{nag05}
Nagaoka,~A.; Watanabe,~N.; Kouchi,~A. H-D Substitution in Interstellar Solid
  Methanol: A Key Route for D Enrichment. \emph{Astrophys. J.} \textbf{2005},
  \emph{L29}, 624\relax
\mciteBstWouldAddEndPuncttrue
\mciteSetBstMidEndSepPunct{\mcitedefaultmidpunct}
{\mcitedefaultendpunct}{\mcitedefaultseppunct}\relax
\EndOfBibitem
\bibitem[Nagaoka \latin{et~al.}(2006)Nagaoka, Watanabe, and Kouchi]{nag06}
Nagaoka,~A.; Watanabe,~N.; Kouchi,~A. Efficient Formation of Deuterated
  Methanol by {H-D} Substitution on Interstellar Grain Surfaces. \emph{AIP
  Conf. Proc.} \textbf{2006}, \emph{855}, 69--75\relax
\mciteBstWouldAddEndPuncttrue
\mciteSetBstMidEndSepPunct{\mcitedefaultmidpunct}
{\mcitedefaultendpunct}{\mcitedefaultseppunct}\relax
\EndOfBibitem
\bibitem[Hiraoka \latin{et~al.}(1994)Hiraoka, Ohashi, Kihara, Yamamoto, Sato,
  and Yamashita]{hir94}
Hiraoka,~K.; Ohashi,~N.; Kihara,~Y.; Yamamoto,~K.; Sato,~T.; Yamashita,~A.
  Formation of Formaldehyde and Methanol from the Reactions of H Atoms with
  Solid CO at 10--20~K. \emph{Chem. Phys. Lett.} \textbf{1994}, \emph{229},
  408--414\relax
\mciteBstWouldAddEndPuncttrue
\mciteSetBstMidEndSepPunct{\mcitedefaultmidpunct}
{\mcitedefaultendpunct}{\mcitedefaultseppunct}\relax
\EndOfBibitem
\bibitem[Watanabe \latin{et~al.}(2003)Watanabe, Shiraki, and Kouchi]{wat03}
Watanabe,~N.; Shiraki,~T.; Kouchi,~A. The Dependence of H$_2$CO and {CH}$_3$OH
  Formation on the Temperature and Thickness of H$_2$O-{CO} Ice during the
  Successive Hydrogenation of {CO}. \emph{Astrophys. J.} \textbf{2003},
  \emph{588}, L121--L124\relax
\mciteBstWouldAddEndPuncttrue
\mciteSetBstMidEndSepPunct{\mcitedefaultmidpunct}
{\mcitedefaultendpunct}{\mcitedefaultseppunct}\relax
\EndOfBibitem
\bibitem[Watanabe \latin{et~al.}(2004)Watanabe, Nagaoka, Shiraki, and
  Kouchi]{wat04}
Watanabe,~N.; Nagaoka,~A.; Shiraki,~T.; Kouchi,~A. Hydrogenation of {CO} on
  Pure Solid {CO} and {CO}-H$_2$O Mixed Ice. \emph{Astrophys. J.}
  \textbf{2004}, \emph{616}, 638--642\relax
\mciteBstWouldAddEndPuncttrue
\mciteSetBstMidEndSepPunct{\mcitedefaultmidpunct}
{\mcitedefaultendpunct}{\mcitedefaultseppunct}\relax
\EndOfBibitem
\bibitem[Garrod \latin{et~al.}(2007)Garrod, Wakelam, and Herbst]{gar07}
Garrod,~R.~T.; Wakelam,~V.; Herbst,~E. Non-Thermal Desorption from Interstellar
  Dust Grains via Exothermic Surface Reactions. \emph{Astron. Astrophys.}
  \textbf{2007}, \emph{467}, 1103--1115\relax
\mciteBstWouldAddEndPuncttrue
\mciteSetBstMidEndSepPunct{\mcitedefaultmidpunct}
{\mcitedefaultendpunct}{\mcitedefaultseppunct}\relax
\EndOfBibitem
\bibitem[Fuchs \latin{et~al.}(2009)Fuchs, Cuppen, Ioppolo, Romanzin, Bisschop,
  Andersson, van Dishoeck, and Linnartz]{fuc09}
Fuchs,~G.~W.; Cuppen,~H.~M.; Ioppolo,~S.; Romanzin,~C.; Bisschop,~S.~E.;
  Andersson,~S.; van Dishoeck,~E.~F.; Linnartz,~H. Hydrogenation Reactions in
  Interstellar CO~Ice Analogues - A Combined Experimental/Theoretical Approach.
  \emph{Astron. Astrophys.} \textbf{2009}, \emph{505}, 629--639\relax
\mciteBstWouldAddEndPuncttrue
\mciteSetBstMidEndSepPunct{\mcitedefaultmidpunct}
{\mcitedefaultendpunct}{\mcitedefaultseppunct}\relax
\EndOfBibitem
\bibitem[Pirim \latin{et~al.}(2010)Pirim, Krim, Laffon, Parent, Pauzat,
  Pilm{\'e}, and Ellinger]{pir10}
Pirim,~C.; Krim,~L.; Laffon,~C.; Parent,~P.; Pauzat,~F.; Pilm{\'e},~J.;
  Ellinger,~Y. Preliminary Study of the Influence of Environment Conditions on
  the Successive Hydrogenations of CO. \emph{J. Phys. Chem. A} \textbf{2010},
  \emph{114}, 3320--3328\relax
\mciteBstWouldAddEndPuncttrue
\mciteSetBstMidEndSepPunct{\mcitedefaultmidpunct}
{\mcitedefaultendpunct}{\mcitedefaultseppunct}\relax
\EndOfBibitem
\bibitem[Morisset \latin{et~al.}(2019)Morisset, Rougeau, and
  Teillet-Billy]{mor19}
Morisset,~S.; Rougeau,~N.; Teillet-Billy,~D. Hydrogenation Reactions and
  Adsorption: From CO to Methanol on a Graphene Surface. \emph{Mol. Astrophys.}
  \textbf{2019}, \emph{14}, 1 -- 9\relax
\mciteBstWouldAddEndPuncttrue
\mciteSetBstMidEndSepPunct{\mcitedefaultmidpunct}
{\mcitedefaultendpunct}{\mcitedefaultseppunct}\relax
\EndOfBibitem
\bibitem[Hidaka \latin{et~al.}(2007)Hidaka, Kouchi, and Watanabe]{hid07}
Hidaka,~H.; Kouchi,~A.; Watanabe,~N. Temperature, Composition, and Hydrogen
  Isotope Effect in the Hydrogenation of CO on Amorphous Ice Surface at
  10--20~K. \emph{J. Chem. Phys.} \textbf{2007}, \emph{126}, 204707\relax
\mciteBstWouldAddEndPuncttrue
\mciteSetBstMidEndSepPunct{\mcitedefaultmidpunct}
{\mcitedefaultendpunct}{\mcitedefaultseppunct}\relax
\EndOfBibitem
\bibitem[Watanabe and Kouchi(2008)Watanabe, and Kouchi]{wat08}
Watanabe,~N.; Kouchi,~A. Ice Surface Reactions: A Key to Chemical Evolution in
  Space. \emph{Prog. Surf. Sci.} \textbf{2008}, \emph{83}, 439--489\relax
\mciteBstWouldAddEndPuncttrue
\mciteSetBstMidEndSepPunct{\mcitedefaultmidpunct}
{\mcitedefaultendpunct}{\mcitedefaultseppunct}\relax
\EndOfBibitem
\bibitem[Andersson \latin{et~al.}(2011)Andersson, Goumans, and
  Arnaldsson]{and11}
Andersson,~S.; Goumans,~T. P.~M.; Arnaldsson,~A. Tunneling in Hydrogen and
  Deuterium Atom Addition to {CO} at Low Temperatures. \emph{Chem. Phys. Lett.}
  \textbf{2011}, \emph{513}, 31--36\relax
\mciteBstWouldAddEndPuncttrue
\mciteSetBstMidEndSepPunct{\mcitedefaultmidpunct}
{\mcitedefaultendpunct}{\mcitedefaultseppunct}\relax
\EndOfBibitem
\bibitem[Song and K{\"a}stner(2017)Song, and K{\"a}stner]{son17}
Song,~L.; K{\"a}stner,~J. Tunneling Rate Constants for H$_{2}$CO + H on
  Amorphous Solid Water Surfaces. \emph{Astrophys. J.} \textbf{2017},
  \emph{850}, 118\relax
\mciteBstWouldAddEndPuncttrue
\mciteSetBstMidEndSepPunct{\mcitedefaultmidpunct}
{\mcitedefaultendpunct}{\mcitedefaultseppunct}\relax
\EndOfBibitem
\bibitem[Goumans and K{\"a}stner(2011)Goumans, and K{\"a}stner]{gou11a}
Goumans,~T. P.~M.; K{\"a}stner,~J. Deuterium Enrichment of Interstellar
  Methanol Explained by Atom Tunneling. \emph{J. Phys. Chem. A} \textbf{2011},
  \emph{115}, 10767\relax
\mciteBstWouldAddEndPuncttrue
\mciteSetBstMidEndSepPunct{\mcitedefaultmidpunct}
{\mcitedefaultendpunct}{\mcitedefaultseppunct}\relax
\EndOfBibitem
\bibitem[Lendvay \latin{et~al.}(1997)Lendvay, B{\'e}rces, and M{\'a}rta]{len97}
Lendvay,~G.; B{\'e}rces,~T.; M{\'a}rta,~F. An ab Initio Study of the
  Three-Channel Reaction between Methanol and Hydrogen Atoms: BAC-MP4 and
  Gaussian-2 Calculations. \emph{J. Phys. Chem. A} \textbf{1997}, \emph{101},
  1588--1594\relax
\mciteBstWouldAddEndPuncttrue
\mciteSetBstMidEndSepPunct{\mcitedefaultmidpunct}
{\mcitedefaultendpunct}{\mcitedefaultseppunct}\relax
\EndOfBibitem
\bibitem[Jodkowski \latin{et~al.}(1999)Jodkowski, Rayez, Rayez, B{\'e}rces, and
  D{\'o}b{\'e}]{jod99}
Jodkowski,~J.~T.; Rayez,~M.-T.; Rayez,~J.-C.; B{\'e}rces,~T.; D{\'o}b{\'e},~S.
  Theoretical Study of the Kinetics of the Hydrogen Abstraction from Methanol.
  3. Reaction of Methanol with Hydrogen Atom, Methyl, and Hydroxyl Radicals.
  \emph{J. Phys. Chem. A} \textbf{1999}, \emph{103}, 3750--3765\relax
\mciteBstWouldAddEndPuncttrue
\mciteSetBstMidEndSepPunct{\mcitedefaultmidpunct}
{\mcitedefaultendpunct}{\mcitedefaultseppunct}\relax
\EndOfBibitem
\bibitem[Kerkeni and Clary(2004)Kerkeni, and Clary]{ker04}
Kerkeni,~B.; Clary,~D.~C. Ab Initio Rate Constants from Hyperspherical Quantum
  Scattering: Application to H+C$_2$H$_6$ and H+CH$_3$OH. \emph{J. Chem. Phys.}
  \textbf{2004}, \emph{121}, 6809--6821\relax
\mciteBstWouldAddEndPuncttrue
\mciteSetBstMidEndSepPunct{\mcitedefaultmidpunct}
{\mcitedefaultendpunct}{\mcitedefaultseppunct}\relax
\EndOfBibitem
\bibitem[Meana-Pa{\~n}eda \latin{et~al.}(2011)Meana-Pa{\~n}eda, Truhlar, and
  Fern{\'a}ndez-Ramos]{mea11}
Meana-Pa{\~n}eda,~R.; Truhlar,~D.~G.; Fern{\'a}ndez-Ramos,~A. High-Level
  Direct-Dynamics Variational Transition State Theory Calculations Including
  Multidimensional Tunneling of the Thermal Rate Constants, Branching Ratios,
  and Kinetic Isotope Effects of the Hydrogen Abstraction Reactions from
  Methanol by Atomic Hydrogen. \emph{J. Chem. Phys.} \textbf{2011}, \emph{134},
  094302\relax
\mciteBstWouldAddEndPuncttrue
\mciteSetBstMidEndSepPunct{\mcitedefaultmidpunct}
{\mcitedefaultendpunct}{\mcitedefaultseppunct}\relax
\EndOfBibitem
\bibitem[Wang and Bowie(2012)Wang, and Bowie]{wan12}
Wang,~T.; Bowie,~J.~H. Hydrogen Tunnelling Influences the Isomerisation of Some
  Small Radicals of Interstellar Importance. A Theoretical Investigation.
  \emph{Org. Biomol. Chem.} \textbf{2012}, \emph{10}, 3219--3228\relax
\mciteBstWouldAddEndPuncttrue
\mciteSetBstMidEndSepPunct{\mcitedefaultmidpunct}
{\mcitedefaultendpunct}{\mcitedefaultseppunct}\relax
\EndOfBibitem
\bibitem[Ryazanov \latin{et~al.}(2012)Ryazanov, Rodrigo, and Reisler]{rya12}
Ryazanov,~M.; Rodrigo,~C.; Reisler,~H. Overtone-Induced Dissociation and
  Isomerization Dynamics of the Hydroxymethyl Radical (CH$_2$OH and CD$_2$OH).
  II. Velocity Map Imaging Studies. \emph{J. Chem. Phys.} \textbf{2012},
  \emph{136}, 084305\relax
\mciteBstWouldAddEndPuncttrue
\mciteSetBstMidEndSepPunct{\mcitedefaultmidpunct}
{\mcitedefaultendpunct}{\mcitedefaultseppunct}\relax
\EndOfBibitem
\bibitem[Meagher \latin{et~al.}(1974)Meagher, Kim, Lee, and Timmons]{mea74}
Meagher,~J.~F.; Kim,~P.; Lee,~J.~H.; Timmons,~R.~B. Kinetic Isotope Effects in
  the Reactions of Hydrogen and Deuterium Atoms with Dimethyl Ether and
  Methanol. \emph{J. Phys. Chem.} \textbf{1974}, \emph{78}, 2650--2657\relax
\mciteBstWouldAddEndPuncttrue
\mciteSetBstMidEndSepPunct{\mcitedefaultmidpunct}
{\mcitedefaultendpunct}{\mcitedefaultseppunct}\relax
\EndOfBibitem
\bibitem[Baulch \latin{et~al.}(2005)Baulch, Bowman, Cobos, Cox, Just, Kerr,
  Pilling, Stocker, Troe, Tsang, Walker, and Warnatz]{bau05}
Baulch,~D.~L.; Bowman,~C.~T.; Cobos,~C.~J.; Cox,~R.~A.; Just,~T.; Kerr,~J.~A.;
  Pilling,~M.~J.; Stocker,~D.; Troe,~J.; Tsang,~W. \latin{et~al.}  Evaluated
  Kinetic Data for Combustion Modeling: Supplement II. \emph{J. Phys. Chem.
  Ref. Data} \textbf{2005}, \emph{34}, 757--1397\relax
\mciteBstWouldAddEndPuncttrue
\mciteSetBstMidEndSepPunct{\mcitedefaultmidpunct}
{\mcitedefaultendpunct}{\mcitedefaultseppunct}\relax
\EndOfBibitem
\bibitem[Carvalho \latin{et~al.}(2008)Carvalho, Barauna, Machado, and
  Roberto-Neto]{car08}
Carvalho,~E.; Barauna,~A.~N.; Machado,~F.~B.; Roberto-Neto,~O. Theoretical
  Calculations of Energetics, Structures, and Rate Constants for the H+CH$_3$OH
  Hydrogen Abstraction Reactions. \emph{Chem. Phys. Lett.} \textbf{2008},
  \emph{463}, 33 -- 37\relax
\mciteBstWouldAddEndPuncttrue
\mciteSetBstMidEndSepPunct{\mcitedefaultmidpunct}
{\mcitedefaultendpunct}{\mcitedefaultseppunct}\relax
\EndOfBibitem
\bibitem[Sanches-Neto \latin{et~al.}(2017)Sanches-Neto, Coutinho, and
  Carvalho-Silva]{san17}
Sanches-Neto,~F.~O.; Coutinho,~N.~D.; Carvalho-Silva,~V.~H. A Novel Assessment
  of the Role of the Methyl Radical and Water Formation Channel in the CH$_3$OH
  + H Reaction. \emph{Phys. Chem. Chem. Phys.} \textbf{2017}, \emph{19},
  24467--24477\relax
\mciteBstWouldAddEndPuncttrue
\mciteSetBstMidEndSepPunct{\mcitedefaultmidpunct}
{\mcitedefaultendpunct}{\mcitedefaultseppunct}\relax
\EndOfBibitem
\bibitem[Shan and Clary(2018)Shan, and Clary]{sha18}
Shan,~X.; Clary,~D.~C. Application of One-Dimensional Semiclassical Transition
  State Theory to the CH$_3$OH + H $\rightarrow$; CH$_2$OH/CH$_3$O + H$_2$
  reactions. \emph{Philos. Trans. Royal Soc. A} \textbf{2018}, \emph{376},
  20170147\relax
\mciteBstWouldAddEndPuncttrue
\mciteSetBstMidEndSepPunct{\mcitedefaultmidpunct}
{\mcitedefaultendpunct}{\mcitedefaultseppunct}\relax
\EndOfBibitem
\bibitem[Cooper \latin{et~al.}(2018)Cooper, Hallmen, and K{\"a}stner]{coo18}
Cooper,~A.~M.; Hallmen,~P.~P.; K{\"a}stner,~J. Potential Energy Surface
  Interpolation with Neural Networks for Instanton Rate Calculations. \emph{J.
  Chem. Phys.} \textbf{2018}, \emph{148}, 094106\relax
\mciteBstWouldAddEndPuncttrue
\mciteSetBstMidEndSepPunct{\mcitedefaultmidpunct}
{\mcitedefaultendpunct}{\mcitedefaultseppunct}\relax
\EndOfBibitem
\bibitem[Meisner \latin{et~al.}(2017)Meisner, Lamberts, and K{\"a}stner]{mei17}
Meisner,~J.; Lamberts,~T.; K{\"a}stner,~J. Atom Tunneling in the Water
  Formation Reaction H$_2$ + OH $\rightarrow$ H$_2$O + H on an Ice Surface.
  \emph{ACS Earth Space Chem.} \textbf{2017}, \emph{1}, 399--410\relax
\mciteBstWouldAddEndPuncttrue
\mciteSetBstMidEndSepPunct{\mcitedefaultmidpunct}
{\mcitedefaultendpunct}{\mcitedefaultseppunct}\relax
\EndOfBibitem
\bibitem[Lamberts \latin{et~al.}(2016)Lamberts, Samanta, K\"ohn, and
  K\"astner]{lam16}
Lamberts,~T.; Samanta,~P.~K.; K\"ohn,~A.; K\"astner,~J. Quantum Tunneling
  During Interstellar Surface-Catalyzed Formation of Water: the Reaction {H +
  H$_2$O$_2$ $\rightarrow$ H$_2$O + OH}. \emph{Phys. Chem. Chem. Phys.}
  \textbf{2016}, \emph{18}, 33021--33030\relax
\mciteBstWouldAddEndPuncttrue
\mciteSetBstMidEndSepPunct{\mcitedefaultmidpunct}
{\mcitedefaultendpunct}{\mcitedefaultseppunct}\relax
\EndOfBibitem
\bibitem[Song and K\"astner(2016)Song, and K\"astner]{son16}
Song,~L.; K\"astner,~J. Formation of the Prebiotic Molecule NH$_2$CHO on
  Astronomical Amorphous Solid Water Surfaces: Accurate Tunneling Rate
  Calculations. \emph{Phys. Chem. Chem. Phys.} \textbf{2016}, \emph{18},
  29278--29285\relax
\mciteBstWouldAddEndPuncttrue
\mciteSetBstMidEndSepPunct{\mcitedefaultmidpunct}
{\mcitedefaultendpunct}{\mcitedefaultseppunct}\relax
\EndOfBibitem
\bibitem[Lamberts and K\"astner(2017)Lamberts, and K\"astner]{lam17a}
Lamberts,~T.; K\"astner,~J. Influence of Surface and Bulk Water Ice on the
  Reactivity of a Water-Forming Reaction. \emph{Astrophys. J.} \textbf{2017},
  \emph{846}, 43\relax
\mciteBstWouldAddEndPuncttrue
\mciteSetBstMidEndSepPunct{\mcitedefaultmidpunct}
{\mcitedefaultendpunct}{\mcitedefaultseppunct}\relax
\EndOfBibitem
\bibitem[Werner \latin{et~al.}(2012)Werner, Knowles, Knizia, Manby,
  {Sch\"{u}tz}, Celani, Korona, Lindh, Mitrushenkov, Rauhut, Shamasundar,
  Adler, Amos, Bernhardsson, Berning, Cooper, Deegan, Dobbyn, Eckert, Goll,
  Hampel, Hesselmann, Hetzer, Hrenar, Jansen, K\"oppl, Liu, Lloyd, Mata, May,
  McNicholas, Meyer, Mura, Nicklass, O'Neill, Palmieri, Peng, Pfl\"uger,
  Pitzer, Reiher, Shiozaki, Stoll, Stone, Tarroni, Thorsteinsson, and
  Wang]{wer12}
Werner,~H.-J.; Knowles,~P.~J.; Knizia,~G.; Manby,~F.~R.; {Sch\"{u}tz},~M.;
  Celani,~P.; Korona,~T.; Lindh,~R.; Mitrushenkov,~A.; Rauhut,~G.
  \latin{et~al.}  MOLPRO, Version 2012.1, a Package of ab Initio Programs.
  2012; see http://www.molpro.net\relax
\mciteBstWouldAddEndPuncttrue
\mciteSetBstMidEndSepPunct{\mcitedefaultmidpunct}
{\mcitedefaultendpunct}{\mcitedefaultseppunct}\relax
\EndOfBibitem
\bibitem[Sherwood \latin{et~al.}(2003)Sherwood, de~Vries, Guest, Schreckenbach,
  Catlow, French, Sokol, Bromley, Thiel, Turner, Billeter, Terstegen, Thiel,
  Kendrick, Rogers, Casci, Watson, King, Karlsen, Sj{\o}voll, Fahmi,
  Sch{\"a}fer, and Lennartz]{she03}
Sherwood,~P.; de~Vries,~A.~H.; Guest,~M.~F.; Schreckenbach,~G.; Catlow,~C.
  R.~A.; French,~S.~A.; Sokol,~A.~A.; Bromley,~S.~T.; Thiel,~W.; Turner,~A.~J.
  \latin{et~al.}  QUASI: A General Purpose Implementation of the QM/MM Approach
  and its Application to Problems in Catalysis. \emph{J. Mol. Struct.
  (THEOCHEM)} \textbf{2003}, \emph{632}, 1\relax
\mciteBstWouldAddEndPuncttrue
\mciteSetBstMidEndSepPunct{\mcitedefaultmidpunct}
{\mcitedefaultendpunct}{\mcitedefaultseppunct}\relax
\EndOfBibitem
\bibitem[Metz \latin{et~al.}(2014)Metz, K{\"a}stner, Sokol, Keal, and
  Sherwood]{met14}
Metz,~S.; K{\"a}stner,~J.; Sokol,~A.~A.; Keal,~T.~W.; Sherwood,~P.
  ChemShell---a Modular Software Package for QM/MM Simulations. \emph{WIREs
  Comput. Mol. Sci.} \textbf{2014}, \emph{4}, 101\relax
\mciteBstWouldAddEndPuncttrue
\mciteSetBstMidEndSepPunct{\mcitedefaultmidpunct}
{\mcitedefaultendpunct}{\mcitedefaultseppunct}\relax
\EndOfBibitem
\bibitem[K{\"a}stner \latin{et~al.}(2009)K{\"a}stner, Carr, Keal, Thiel,
  Wander, and Sherwood]{kae09}
K{\"a}stner,~J.; Carr,~J.~M.; Keal,~T.~W.; Thiel,~W.; Wander,~A.; Sherwood,~P.
  DL-FIND: An Open-Source Geometry Optimizer for Atomistic Simulations.
  \emph{J. Phys. Chem. A} \textbf{2009}, \emph{113}, 11856--11865\relax
\mciteBstWouldAddEndPuncttrue
\mciteSetBstMidEndSepPunct{\mcitedefaultmidpunct}
{\mcitedefaultendpunct}{\mcitedefaultseppunct}\relax
\EndOfBibitem
\bibitem[Liu and Nocedal(1989)Liu, and Nocedal]{liu89}
Liu,~D.~C.; Nocedal,~J. On the Limited Memory BFGS Method for Large Scale
  Optimization. \emph{Math. Prog.} \textbf{1989}, \emph{45}, 503--528\relax
\mciteBstWouldAddEndPuncttrue
\mciteSetBstMidEndSepPunct{\mcitedefaultmidpunct}
{\mcitedefaultendpunct}{\mcitedefaultseppunct}\relax
\EndOfBibitem
\bibitem[Langer(1967)]{lan67}
Langer,~J.~S. Theory of the Condensation Point. \emph{Ann. Phys. (N.Y.)}
  \textbf{1967}, \emph{41}, 108\relax
\mciteBstWouldAddEndPuncttrue
\mciteSetBstMidEndSepPunct{\mcitedefaultmidpunct}
{\mcitedefaultendpunct}{\mcitedefaultseppunct}\relax
\EndOfBibitem
\bibitem[Miller(1975)]{mil75}
Miller,~W.~H. Semiclassical Limit of Quantum Mechanical Transition State Theory
  for Nonseparable Systems. \emph{J. Chem. Phys.} \textbf{1975}, \emph{62},
  1899\relax
\mciteBstWouldAddEndPuncttrue
\mciteSetBstMidEndSepPunct{\mcitedefaultmidpunct}
{\mcitedefaultendpunct}{\mcitedefaultseppunct}\relax
\EndOfBibitem
\bibitem[Coleman(1977)]{col77}
Coleman,~S. Fate of the False Vacuum: Semiclassical Theory. \emph{Phys. Rev. D}
  \textbf{1977}, \emph{15}, 2929\relax
\mciteBstWouldAddEndPuncttrue
\mciteSetBstMidEndSepPunct{\mcitedefaultmidpunct}
{\mcitedefaultendpunct}{\mcitedefaultseppunct}\relax
\EndOfBibitem
\bibitem[{Callan Jr.} and Coleman(1977){Callan Jr.}, and Coleman]{cal77}
{Callan Jr.},~C.~G.; Coleman,~S. Fate of the False Vacuum. II. First Quantum
  Corrections. \emph{Phys. Rev. D} \textbf{1977}, \emph{16}, 1762\relax
\mciteBstWouldAddEndPuncttrue
\mciteSetBstMidEndSepPunct{\mcitedefaultmidpunct}
{\mcitedefaultendpunct}{\mcitedefaultseppunct}\relax
\EndOfBibitem
\bibitem[Richardson and Althorpe(2009)Richardson, and Althorpe]{ric09}
Richardson,~J.~O.; Althorpe,~S.~C. Ring-Polymer Molecular Dynamics Rate-Theory
  in the Deep-Tunneling Regime: Connection with Semiclassical Instanton Theory.
  \emph{J. Chem. Phys.} \textbf{2009}, \emph{131}, 214106\relax
\mciteBstWouldAddEndPuncttrue
\mciteSetBstMidEndSepPunct{\mcitedefaultmidpunct}
{\mcitedefaultendpunct}{\mcitedefaultseppunct}\relax
\EndOfBibitem
\bibitem[Althorpe(2011)]{alt11}
Althorpe,~S.~C. On the Equivalence of Two Commonly Used Forms of Semiclassical
  Instanton Theory. \emph{J. Chem. Phys.} \textbf{2011}, \emph{134},
  114104\relax
\mciteBstWouldAddEndPuncttrue
\mciteSetBstMidEndSepPunct{\mcitedefaultmidpunct}
{\mcitedefaultendpunct}{\mcitedefaultseppunct}\relax
\EndOfBibitem
\bibitem[K{\"a}stner(2014)]{kae14}
K{\"a}stner,~J. Theory and Simulation of Atom Tunneling in Chemical Reactions.
  \emph{Wiley Interdiscip. Rev.-Comput. Mol. Sci} \textbf{2014}, \emph{4},
  158--168\relax
\mciteBstWouldAddEndPuncttrue
\mciteSetBstMidEndSepPunct{\mcitedefaultmidpunct}
{\mcitedefaultendpunct}{\mcitedefaultseppunct}\relax
\EndOfBibitem
\bibitem[Richardson(2016)]{ric16}
Richardson,~J.~O. Derivation of Instanton Rate Theory from First Principles.
  \emph{J. Chem. Phys.} \textbf{2016}, \emph{144}, 114106\relax
\mciteBstWouldAddEndPuncttrue
\mciteSetBstMidEndSepPunct{\mcitedefaultmidpunct}
{\mcitedefaultendpunct}{\mcitedefaultseppunct}\relax
\EndOfBibitem
\bibitem[Rommel \latin{et~al.}(2011)Rommel, Goumans, and K{\"a}stner]{rom11}
Rommel,~J.~B.; Goumans,~T. P.~M.; K{\"a}stner,~J. Locating Instantons in Many
  Degrees of Freedom. \emph{J. Chem. Theory Comput.} \textbf{2011}, \emph{7},
  690--698\relax
\mciteBstWouldAddEndPuncttrue
\mciteSetBstMidEndSepPunct{\mcitedefaultmidpunct}
{\mcitedefaultendpunct}{\mcitedefaultseppunct}\relax
\EndOfBibitem
\bibitem[Rommel and K{\"a}stner(2011)Rommel, and K{\"a}stner]{rom11b}
Rommel,~J.~B.; K{\"a}stner,~J. Adaptive Integration Grids in Instanton Theory
  Improve the Numerical Accuracy at Low Temperature. \emph{J. Chem. Phys.}
  \textbf{2011}, \emph{134}, 184107\relax
\mciteBstWouldAddEndPuncttrue
\mciteSetBstMidEndSepPunct{\mcitedefaultmidpunct}
{\mcitedefaultendpunct}{\mcitedefaultseppunct}\relax
\EndOfBibitem
\bibitem[Kryvohuz(2011)]{kry11}
Kryvohuz,~M. Semiclassical Instanton Approach to Calculation of Reaction Rate
  Constants in Multidimensional Chemical Systems. \emph{J. Chem. Phys.}
  \textbf{2011}, \emph{134}, 114103\relax
\mciteBstWouldAddEndPuncttrue
\mciteSetBstMidEndSepPunct{\mcitedefaultmidpunct}
{\mcitedefaultendpunct}{\mcitedefaultseppunct}\relax
\EndOfBibitem
\bibitem[McConnell and K{\"a}stner(2017)McConnell, and K{\"a}stner]{mcc17a}
McConnell,~S.; K{\"a}stner,~J. Instanton Rate Constant Calculations Close To
  and Above the Crossover Temperature. \emph{J. Comput. Chem.} \textbf{2017},
  \emph{38}, 2570--2580\relax
\mciteBstWouldAddEndPuncttrue
\mciteSetBstMidEndSepPunct{\mcitedefaultmidpunct}
{\mcitedefaultendpunct}{\mcitedefaultseppunct}\relax
\EndOfBibitem
\bibitem[Fern{\'a}ndez-Ramos \latin{et~al.}(2007)Fern{\'a}ndez-Ramos,
  Ellingson, Meana-Pa{\~n}eda, Marques, and Truhlar]{fer07a}
Fern{\'a}ndez-Ramos,~A.; Ellingson,~B.~A.; Meana-Pa{\~n}eda,~R.; Marques,~J.
  M.~C.; Truhlar,~D.~G. Symmetry Numbers and Chemical Reaction Rates.
  \emph{Theor. Chem. Acc.} \textbf{2007}, \emph{118}, 813--826\relax
\mciteBstWouldAddEndPuncttrue
\mciteSetBstMidEndSepPunct{\mcitedefaultmidpunct}
{\mcitedefaultendpunct}{\mcitedefaultseppunct}\relax
\EndOfBibitem
\bibitem[Richardson(2016)]{ric16a}
Richardson,~J.~O. Microcanonical and Thermal Instanton Rate Theory for Chemical
  Reactions at All Temperatures. \emph{Faraday Disc.} \textbf{2016},
  \emph{195}, 49--67\relax
\mciteBstWouldAddEndPuncttrue
\mciteSetBstMidEndSepPunct{\mcitedefaultmidpunct}
{\mcitedefaultendpunct}{\mcitedefaultseppunct}\relax
\EndOfBibitem
\bibitem[McConnell \latin{et~al.}(2017)McConnell, L{\"o}hle, and
  K{\"a}stner]{mcc17}
McConnell,~S.~R.; L{\"o}hle,~A.; K{\"a}stner,~J. Rate Constants from Instanton
  Theory via a Microcanonical Approach. \emph{J. Chem. Phys.} \textbf{2017},
  \emph{146}, 074105\relax
\mciteBstWouldAddEndPuncttrue
\mciteSetBstMidEndSepPunct{\mcitedefaultmidpunct}
{\mcitedefaultendpunct}{\mcitedefaultseppunct}\relax
\EndOfBibitem
\bibitem[L\"ohle and K\"astner(2018)L\"ohle, and K\"astner]{loh18}
L\"ohle,~A.; K\"astner,~J. Calculation of Reaction Rate Constants in the
  Canonical and Microcanonical Ensemble. \emph{J. Chem. Theory Comput.}
  \textbf{2018}, \emph{14}, 5489--5498\relax
\mciteBstWouldAddEndPuncttrue
\mciteSetBstMidEndSepPunct{\mcitedefaultmidpunct}
{\mcitedefaultendpunct}{\mcitedefaultseppunct}\relax
\EndOfBibitem
\bibitem[Senevirathne \latin{et~al.}(2017)Senevirathne, Andersson, Dulieu, and
  Nyman]{sen17}
Senevirathne,~B.; Andersson,~S.; Dulieu,~F.; Nyman,~G. Hydrogen Atom Mobility,
  Kinetic Isotope Effects and Tunneling on Interstellar Ices ({Ih and ASW}).
  \emph{Mol. Astrophys.} \textbf{2017}, \emph{6}, 59--69\relax
\mciteBstWouldAddEndPuncttrue
\mciteSetBstMidEndSepPunct{\mcitedefaultmidpunct}
{\mcitedefaultendpunct}{\mcitedefaultseppunct}\relax
\EndOfBibitem
\bibitem[{\'A}sgeirsson \latin{et~al.}(2017){\'A}sgeirsson, J{\'o}nsson, and
  Wikfeldt]{asg17}
{\'A}sgeirsson,~V.; J{\'o}nsson,~H.; Wikfeldt,~K.~T. Long-Time Scale
  Simulations of Tunneling-Assisted Diffusion of Hydrogen on Ice Surfaces at
  Low Temperature. \emph{J. Phys. Chem. C} \textbf{2017}, \emph{121},
  1648--1657\relax
\mciteBstWouldAddEndPuncttrue
\mciteSetBstMidEndSepPunct{\mcitedefaultmidpunct}
{\mcitedefaultendpunct}{\mcitedefaultseppunct}\relax
\EndOfBibitem
\bibitem[Hama \latin{et~al.}(2012)Hama, Kuwahata, Watanabe, Kouchi, Kimura,
  Chigai, and Pirronello]{ham12}
Hama,~T.; Kuwahata,~K.; Watanabe,~N.; Kouchi,~A.; Kimura,~Y.; Chigai,~T.;
  Pirronello,~V. The Mechanism of Surface Diffusion of H and D Atoms on
  Amorphous Solid Water: Existence of Various Potential Sites. \emph{Astrophys.
  J.} \textbf{2012}, \emph{757}, 185\relax
\mciteBstWouldAddEndPuncttrue
\mciteSetBstMidEndSepPunct{\mcitedefaultmidpunct}
{\mcitedefaultendpunct}{\mcitedefaultseppunct}\relax
\EndOfBibitem
\bibitem[Hama and Watanabe(2013)Hama, and Watanabe]{ham13}
Hama,~T.; Watanabe,~N. Surface Processes on Interstellar Amorphous Solid Water:
  Adsorption, Diffusion, Tunneling Reactions, and Nuclear-Spin Conversion.
  \emph{Chem. Rev.} \textbf{2013}, \emph{113}, 8783--8839\relax
\mciteBstWouldAddEndPuncttrue
\mciteSetBstMidEndSepPunct{\mcitedefaultmidpunct}
{\mcitedefaultendpunct}{\mcitedefaultseppunct}\relax
\EndOfBibitem
\bibitem[Bizzocchi \latin{et~al.}(2014)Bizzocchi, Caselli, Spezzano, and
  Leonardo]{biz14}
Bizzocchi,~L.; Caselli,~P.; Spezzano,~S.; Leonardo,~E. Deuterated Methanol in
  the Pre-Stellar Core L1544. \emph{Astron. Astrophys.} \textbf{2014},
  \emph{569}, A27\relax
\mciteBstWouldAddEndPuncttrue
\mciteSetBstMidEndSepPunct{\mcitedefaultmidpunct}
{\mcitedefaultendpunct}{\mcitedefaultseppunct}\relax
\EndOfBibitem
\bibitem[Chac{\'o}n-Tanarro \latin{et~al.}(2019)Chac{\'o}n-Tanarro, Caselli,
  Bizzocchi, Pineda, Sipil{\"a}, Vasyunin, Spezzano, Punanova, Giuliano, and
  Lattanzi]{cha19}
Chac{\'o}n-Tanarro,~A.; Caselli,~P.; Bizzocchi,~L.; Pineda,~J.~E.;
  Sipil{\"a},~O.; Vasyunin,~A.; Spezzano,~S.; Punanova,~A.; Giuliano,~B.~M.;
  Lattanzi,~V. Mapping Deuterated Methanol Toward L1544 - I. Deuterium Fraction
  and Comparison with Modeling. \emph{Astron. Astrophys.} \textbf{2019},
  \emph{622}, A141\relax
\mciteBstWouldAddEndPuncttrue
\mciteSetBstMidEndSepPunct{\mcitedefaultmidpunct}
{\mcitedefaultendpunct}{\mcitedefaultseppunct}\relax
\EndOfBibitem
\bibitem[Gao \latin{et~al.}(2018)Gao, Zheng, Fernandez-Ramos, Truhlar, and
  Xu]{gao18}
Gao,~L.~G.; Zheng,~J.; Fernandez-Ramos,~A.; Truhlar,~D.~G.; Xu,~X. Kinetics of
  the Methanol Reaction with OH at Interstellar, Atmospheric, and Combustion
  Temperatures. \emph{J. Am. Chem. Soc.} \textbf{2018}, \emph{140},
  2906--2918\relax
\mciteBstWouldAddEndPuncttrue
\mciteSetBstMidEndSepPunct{\mcitedefaultmidpunct}
{\mcitedefaultendpunct}{\mcitedefaultseppunct}\relax
\EndOfBibitem
\end{mcitethebibliography}

\newpage
\section*{TOC Graphic}
\includegraphics[height=3.5cm]{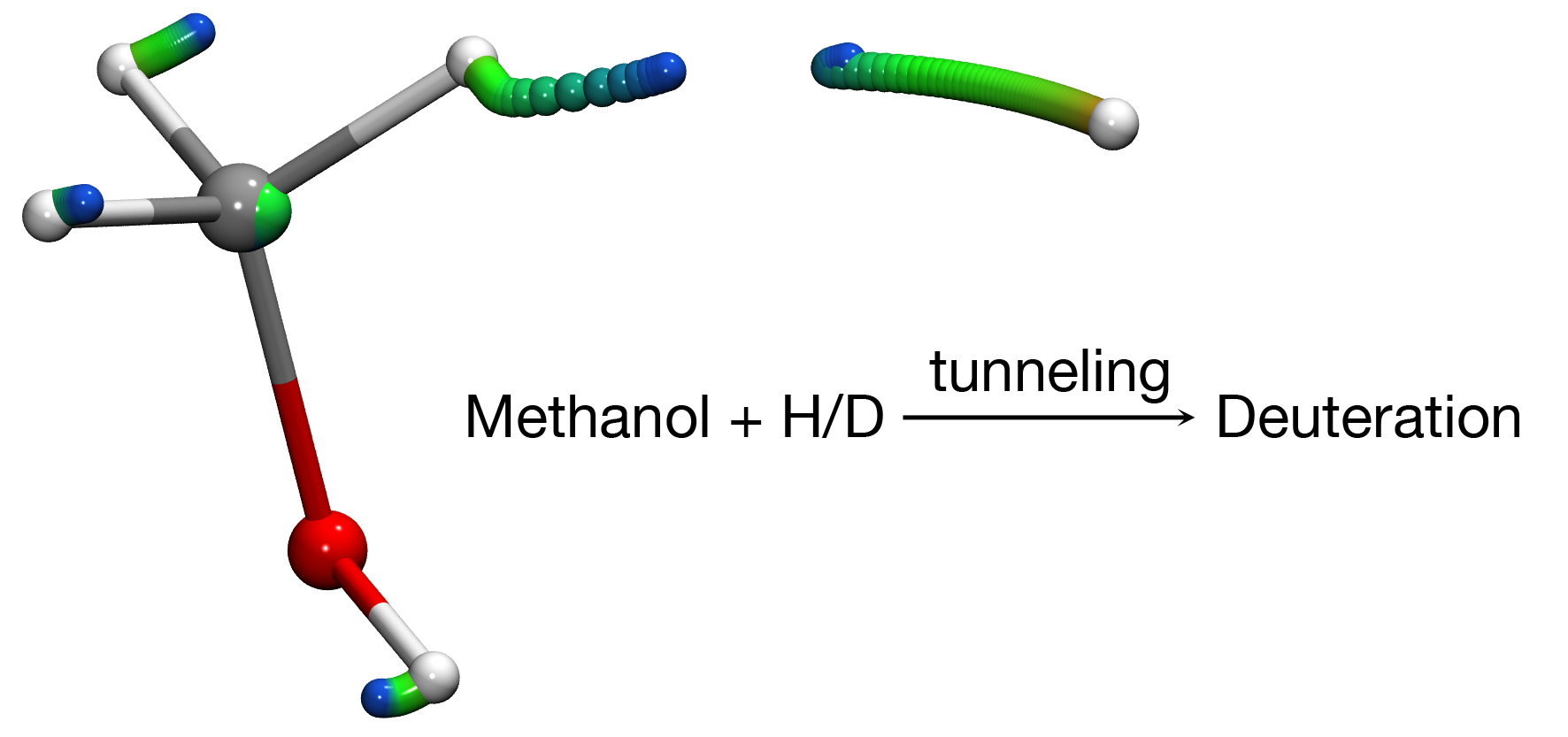}

\end{document}